\documentclass[12pt]{article}

\usepackage[margin=1in]{geometry}
\usepackage[T1]{fontenc}
\usepackage{graphicx}
\usepackage{longtable}
\usepackage{overpic}
\usepackage{amssymb}
\usepackage{amsmath}
\usepackage{tabularx}
\usepackage[numbers,sort&compress]{natbib}

\title{Periodic Trends and a Physics-Based Multistage Search for High-$T_c$ Hydride Superconductors}

\author{
Tianran Chen$^{*,1}$ and Taner Yildirim$^{*,2}$\\[1.2em]
\small $^{1}$Department of Physics and Astronomy, University of Tennessee,\\
\small Knoxville, Tennessee 37996, USA\\
\small $^{2}$NIST Center for Neutron Research, National Institute of Standards and Technology,\\
\small Gaithersburg, Maryland 20899-8562, USA\\[0.8em]
\small $^{*}$To whom correspondence should be addressed:\\
\small tchen34@utk.edu and taner@nist.gov.
}
\date{}

\begin{document} 
\maketitle

\begin{abstract}

Hydrogen-rich materials under pressure are candidates for conventional phonon-mediated superconductivity, but first-principles electron--phonon-coupling (EPC) calculations are computationally demanding. We develop an interpretable multistage workflow for prioritizing binary hydride superconductors before full EPC calculations. Using controlled cubic $X_m$H$_n$ prototypes at 200~GPa, we introduce the spectral moment $A_1$ as an intermediate measure of electron--phonon perturbation strength. In its linewidth-based construction, $A_1$ remains finite as harmonic phonons soften, allowing related structures with imaginary harmonic modes to be screened; it does not establish superconductivity or physical realizability for unstable structures. For dynamically stable cubic systems, $A_1$ correlates strongly with the H-$1s$ density-of-states fraction at the Fermi level, $\alpha$, a projected Fermi-surface descriptor, $\beta$, and atomic number density, $\rho$. These trends are summarized heuristically by $A_1\approx h\alpha\beta\rho$, where $h$ is framework- and pressure-dependent; the descriptor is used for ranking, not as a universal $T_c$ predictor.

We incorporate $\alpha\beta\rho$ and enthalpic competitiveness into a three-stage evolutionary search comprising multiobjective structural screening, harmonic-phonon stability filtering, and coarse density-functional perturbation-theory (DFPT) EPC calculations. Application to more than 100,000 binary hydride structures at 50 and 200~GPa identifies candidates and prioritizes NaH$_6$ structural families. Higher-accuracy calculations show that $Pm\bar{3}m$ NaH$_6$ is dynamically stable down to 50~GPa and has a harmonic Allen--Dynes estimate of $T_c\approx230$~K. The framework combines electronic descriptors with progressively more expensive stability and EPC calculations for scalable hydride exploration.

\end{abstract}

\section*{Introduction and Background}

The search for materials with increasingly high superconducting transition temperatures has remained a central objective of condensed-matter physics since the discovery of superconductivity in mercury. Hydrogen-rich materials under pressure have provided particularly important examples of conventional phonon-mediated superconductivity: H$_3$S and LaH$_{10}$-based systems exhibit transition temperatures above 200~K at megabar pressures \cite{duan2014pressure,errea2015high,liu2017potential,somayazulu2019evidence}. Experimental investigation of these materials remains challenging because of the required pressures and the difficulty of characterizing their high-pressure structures. First-principles calculations therefore play an essential role in identifying candidate hydrides, predicting their structures, and guiding experiments.

Several approaches have been developed to accelerate the computational discovery of superconducting hydrides. High-throughput first-principles searches combined with crystal-structure prediction have enabled systematic exploration of binary hydrides over broad composition and pressure ranges \cite{shipley2021high,saha2023high}. Data-driven and machine-learning approaches have further been used to screen superconducting materials, guide structure searches, and model quantities derived from the Eliashberg spectral function \cite{hutcheon2020predicting,wines2024machine,xie2022machine}. These approaches can access very large search spaces, but their reliability depends on the coverage and consistency of the underlying calculations or training data. Physically interpretable descriptors remain complementary because they can expose trends, guide evolutionary structure searches, and prioritize candidates before expensive electron--phonon coupling (EPC) calculations. To make the relevant computational bottleneck explicit, we briefly summarize the conventional framework used here to estimate $T_c$.

For conventional superconductors, the transition temperature can be estimated from the Eliashberg spectral function $\alpha^2F(\omega)$ using the Allen--Dynes modified McMillan formula \cite{allen1975transition}:
\begin{equation}
T_c=\frac{f_1f_2\omega_{\log}}{1.20}
\exp\left[-\frac{1.04(1+\lambda)}
{\lambda-\mu^*(1+0.62\lambda)}\right],
\label{eq:AllenDynes}
\end{equation}
where
\begin{equation}
\lambda=2\int_0^{\omega_{\max}}\frac{\alpha^2F(\omega)}{\omega}\,d\omega,
\qquad
\omega_{\log}=\exp\left[\frac{2}{\lambda}
\int_0^{\omega_{\max}}\frac{\alpha^2F(\omega)}{\omega}\ln\omega\,d\omega\right],
\end{equation}
and $\mu^*$ is the Coulomb pseudopotential. The strong-coupling
correction factor is
\begin{equation}
f_1=\left[1+\left(\frac{\lambda}{2.46(1+3.8\mu^*)}\right)^{3/2}\right]^{1/3}.
\end{equation}
The spectral-shape correction factor is
\begin{equation}
f_2=1+\frac{\lambda^2(\omega_2/\omega_{\log}-1)}
{\lambda^2+\left[1.82(1+6.3\mu^*)\omega_2/\omega_{\log}\right]^2},
\end{equation}
where
\begin{equation}
\omega_2=\left[\frac{2}{\lambda}
\int_0^{\omega_{\max}}\alpha^2F(\omega)\omega\,d\omega\right]^{1/2}.
\end{equation}
In this work, all reported transition temperatures use $\mu^*=0.10$. The formulae and numerical conventions used to generate the reported $T_c$ values should be checked against the implementation used in the calculations before submission.

Accurate evaluation of $\alpha^2F(\omega)$ requires sufficiently dense sampling of electronic $\mathbf{k}$ points and phonon $\mathbf{q}$ points and can therefore be computationally demanding. In addition, harmonic isotropic Allen--Dynes estimates are screening-level quantities: quantitative predictions for individual hydrides can require denser EPC sampling and, where relevant, anharmonic, quantum-nuclear, anisotropic, or nonadiabatic treatments \cite{errea2015high,pellegrini2024abinitio}. An efficient discovery workflow should consequently use inexpensive quantities to identify candidates for subsequent phonon and EPC calculations, while retaining a transparent connection to the electronic and structural factors that control electron--phonon coupling.

Here, we pursue this strategy by examining systematic trends in a controlled family of binary cubic hydrides. We introduce the electron--perturbation measure $A_1$, a spectral moment that can be evaluated as an intermediate screening quantity even when harmonic phonons are unstable. We then identify empirical correlations between $A_1$ and the hydrogen $1s$ contribution at the Fermi level, the projected Fermi-surface descriptor, and the atomic number density, summarized by the low-cost descriptor $\alpha\beta\rho$. These trends motivate a three-stage workflow in which $\alpha\beta\rho$ and enthalpic competitiveness are used in multiobjective evolutionary searches, harmonic phonons provide a dynamical-stability filter, and coarse DFPT EPC calculations are reserved for the remaining candidates.

The purpose of this work is to establish an interpretable candidate-prioritization framework, rather than a universal predictor of $T_c$. The descriptor relation is calibrated principally for related cubic hydrides at 200~GPa with broadly comparable phonon-frequency scales; outside this domain, a large $\alpha\beta\rho$ identifies a candidate for explicit phonon and EPC calculations rather than establishing a quantitative transition temperature. Applying the workflow to binary hydrides at 50 and 200~GPa identifies several promising structural families, including NaH$_6$, whose detailed structural, vibrational, and superconducting properties will be reported separately.

\section*{The Electron--Perturbation Coupling Constant $A_1$}

The conventional electron--phonon coupling constant $\lambda$ is particularly sensitive to low-frequency phonons because it is weighted by $1/\omega$ in the Eliashberg spectral function. In particular, $\lambda$ is formally divergent if $\alpha^2F(\omega)$ remains nonzero as $\omega\rightarrow0$. Consequently, $\lambda$, $\omega_{\log}$, and the harmonic Allen--Dynes estimate of $T_c$ are not well-defined screening quantities for structures with imaginary or vanishing harmonic phonon frequencies. Such structures occur frequently in crystal-structure searches and may become stable after further relaxation, compression, chemical substitution or doping, anharmonic effects, or transformation to a lower-symmetry phase.

A controlled comparison of compounds within a common structural prototype is useful for identifying trends in superconductivity. However, many elemental substitutions within a given prototype are dynamically unstable at a specified pressure, and the number of dynamically stable hydrogen-rich members of one prototype can be limited. We therefore seek an intermediate quantity that characterizes electron--phonon perturbation strength and can be used to rank related candidate structures before final phonon and EPC calculations. This quantity is a screening descriptor, not a replacement for phonon calculations or a definition of $T_c$ for a dynamically unstable structure.

The Allen--Dynes transition temperature is a nonlinear functional of $\alpha^2F(\omega)$ and cannot be evaluated meaningfully for imaginary harmonic frequencies. We introduce the spectral moments
\begin{equation}
A_n=\int_0^{\omega_{\max}}\omega^n\alpha^2F(\omega)\,d\omega,
\label{eq:An}
\end{equation}
which are linear functionals of $\alpha^2F(\omega)$. By definition, $\lambda=2A_{-1}$. In a linewidth representation, the moments are written as \cite{allen1980dynamical}
\begin{equation}
\gamma_{\mathbf{q}\nu}=\pi N_F\omega_{\mathbf{q}\nu}^2\lambda_{\mathbf{q}\nu},
\qquad
\lambda_{\mathbf{q}\nu}=\frac{1}{N_F\omega_{\mathbf{q}\nu}}
\sum_{mn\mathbf{k}}w_{\mathbf{k}}\left|g_{mn}^{\nu}(\mathbf{k},\mathbf{q})\right|^2
\delta(E_{n\mathbf{k}}-E_F)\delta(E_{m\mathbf{k}+\mathbf{q}}-E_F),
\label{eq:linewidth}
\end{equation}
\begin{equation}
A_{-1}=\frac{1}{2\pi N_F}\sum_{\mathbf{q}\nu}w_{\mathbf{q}}\frac{\gamma_{\mathbf{q}\nu}}{\omega_{\mathbf{q}\nu}^2},
\qquad
A_0=\frac{1}{2\pi N_F}\sum_{\mathbf{q}\nu}w_{\mathbf{q}}\frac{\gamma_{\mathbf{q}\nu}}{\omega_{\mathbf{q}\nu}},
\qquad
A_1=\frac{1}{2\pi N_F}\sum_{\mathbf{q}\nu}w_{\mathbf{q}}\gamma_{\mathbf{q}\nu}.
\label{eq:Anlinewidth}
\end{equation}
Here, $w_{\mathbf{q}}$ and $w_{\mathbf{k}}$ are Brillouin-zone weights, and $N_F$ is the density of states at the Fermi level in the convention used for the linewidth calculation.

\begin{figure}[ht]
\begin{overpic}[width=\textwidth]{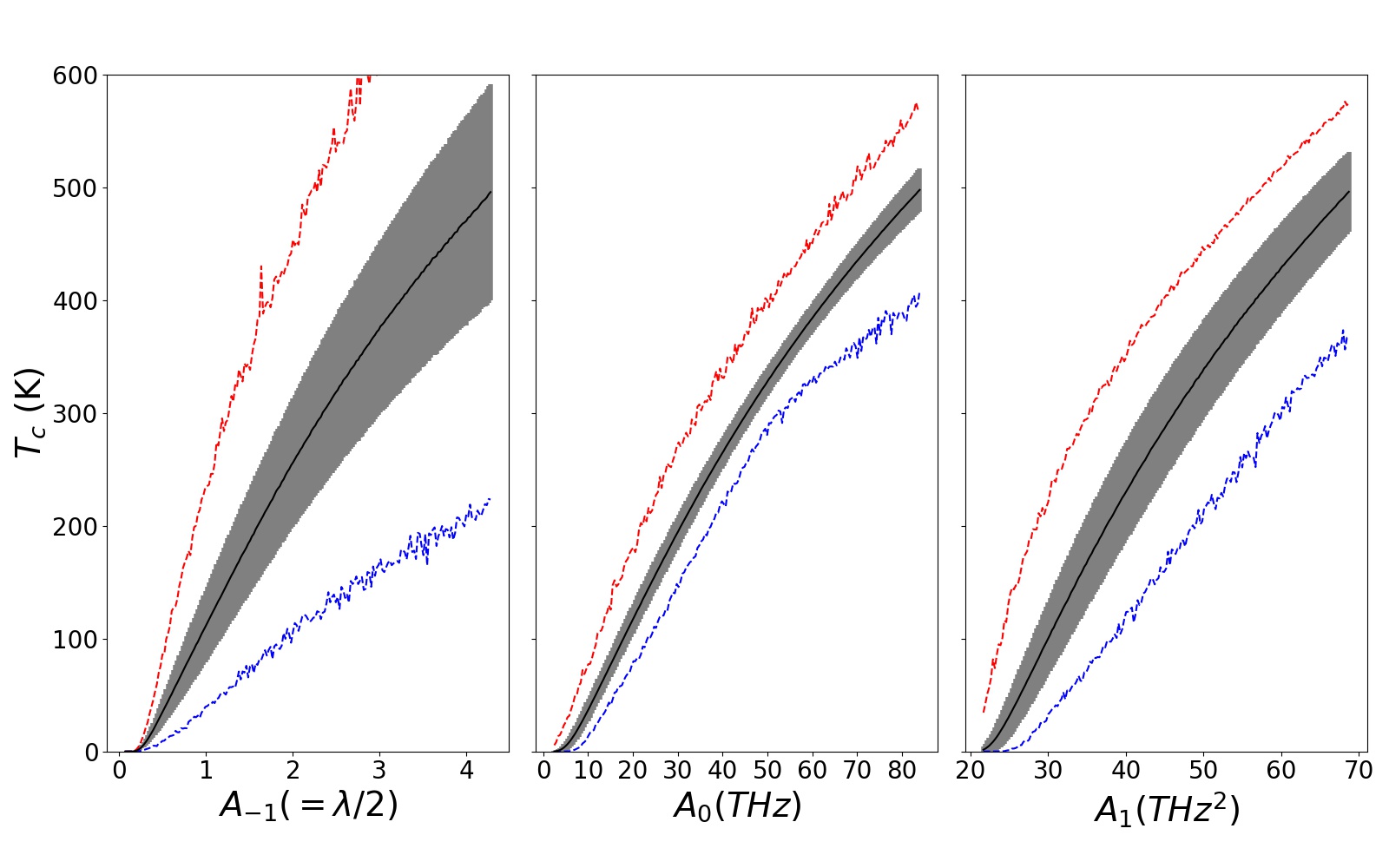}
\put (9,54) {\large\textbf A}
\put (40,54) {\large\textbf B}
\put (71,54) {\large\textbf C}
\end{overpic}
\caption{Mean, standard error, maximum, and minimum values of $T_c$ as functions of the spectral moment $A_n$, obtained from randomly generated $\alpha^2F(\omega)$ spectra: (A) $n=-1$, (B) $n=0$, and (C) $n=1$. Gray shaded regions indicate the standard errors, and red and blue dashed lines denote the maximum and minimum values of $T_c$, respectively.}
\label{fig-Tc_An}
\end{figure}

To assess how strongly the moments constrain $T_c$, we generated random nonnegative $\alpha^2F(\omega)$ spectra subject to $\alpha^2F(0)=0$ and maximum phonon frequencies spanning $\omega_{\max}=57\pm9$~THz, the distribution obtained for the dynamically stable cubic systems analyzed here. For each specified value $A_n=x$, 20,000 spectra were generated and normalized to satisfy this constraint. We then evaluated $T_c$ using Eq.~\eqref{eq:AllenDynes} and determined the mean, standard error, maximum, and minimum values. The results, shown in Fig.~\ref{fig-Tc_An}, demonstrate that $A_0$ provides the strongest correlation with $T_c$ among the three moments considered. Quan \textit{et al.}\ likewise found that $A_0$ is a particularly effective moment for correlating $T_c$ across calculations for H$_3$S, CaH$_6$, MgH$_6$, LaH$_{10}$, and YH$_{10}$ at several pressures \cite{quan2019compressed}. Our use of $A_1$ instead trades some of this stable-system correlation for an intermediate screening quantity that remains finite in the linewidth-based construction when harmonic modes soften.

As indicated by Eq.~\eqref{eq:Anlinewidth}, both $A_{-1}$ and $A_0$ contain inverse powers of the phonon frequency and diverge when a mode with $\omega_{\mathbf{q}\nu}=0$ has nonzero linewidth. They therefore cannot be used to characterize structures with imaginary harmonic phonons; see Appendix~2 for the mode-resolved behavior near a soft mode. In contrast, $A_1$ contains no inverse phonon-frequency factor in its linewidth representation. Starting from
\begin{equation}
\alpha^2F(\omega)=\frac{1}{2}\sum_{\mathbf{q}\nu}w_{\mathbf{q}}\omega_{\mathbf{q}\nu}\lambda_{\mathbf{q}\nu}
\delta(\omega-\omega_{\mathbf{q}\nu}),
\label{eq:a2F}
\end{equation}
and Eq.~\eqref{eq:linewidth}, we obtain
\begin{equation}
\begin{split}
A_1&=\int_0^{\omega_{\max}}\omega\alpha^2F(\omega)\,d\omega\\
&=\frac{1}{2N_F}\sum_{\nu,mn,\mathbf{q}\mathbf{k}}w_{\mathbf{q}}w_{\mathbf{k}}\omega_{\mathbf{q}\nu}
\left|g_{mn}^{\nu}(\mathbf{k},\mathbf{q})\right|^2
\delta(E_{n\mathbf{k}}-E_F)\delta(E_{m\mathbf{k}+\mathbf{q}}-E_F).
\end{split}
\label{eq:A1g}
\end{equation}

Equation~\eqref{eq:A1g} motivates interpreting $A_1$ as an electron--perturbation measure. A formulation in terms of mass-weighted deformation-potential matrix elements can, in principle, remove the explicit frequency factor in Eq.~\eqref{eq:A1g}. For a multicomponent hydride, however, that transformation must be written in the full atom- and Cartesian-resolved mass-weighted displacement basis; a single scalar mass $M$ is not generally sufficient. Moreover, whether the matrix element supplied by a particular \textsc{Quantum ESPRESSO} workflow already contains phonon-normalization factors must be established before such a transformation is used. We therefore use the linewidth-based definition in Eq.~\eqref{eq:Anlinewidth} for the reported values of $A_1$.

Because $A_1$ is less sensitive than $\lambda$, $\omega_{\log}$, and $T_c$ to the detailed phonon spectrum, it may also converge with coarser $\mathbf{q}$- and $\mathbf{k}$-point meshes. We test this expectation for $Fm\bar{3}m$ H$_3$S at 200~GPa using $N_q\times N_q\times N_q$ $\mathbf{q}$-point meshes and $N_k\times N_k\times N_k$ $\mathbf{k}$-point meshes, where $N_k$ is an integer multiple of $N_q$. All other computational parameters are held fixed.

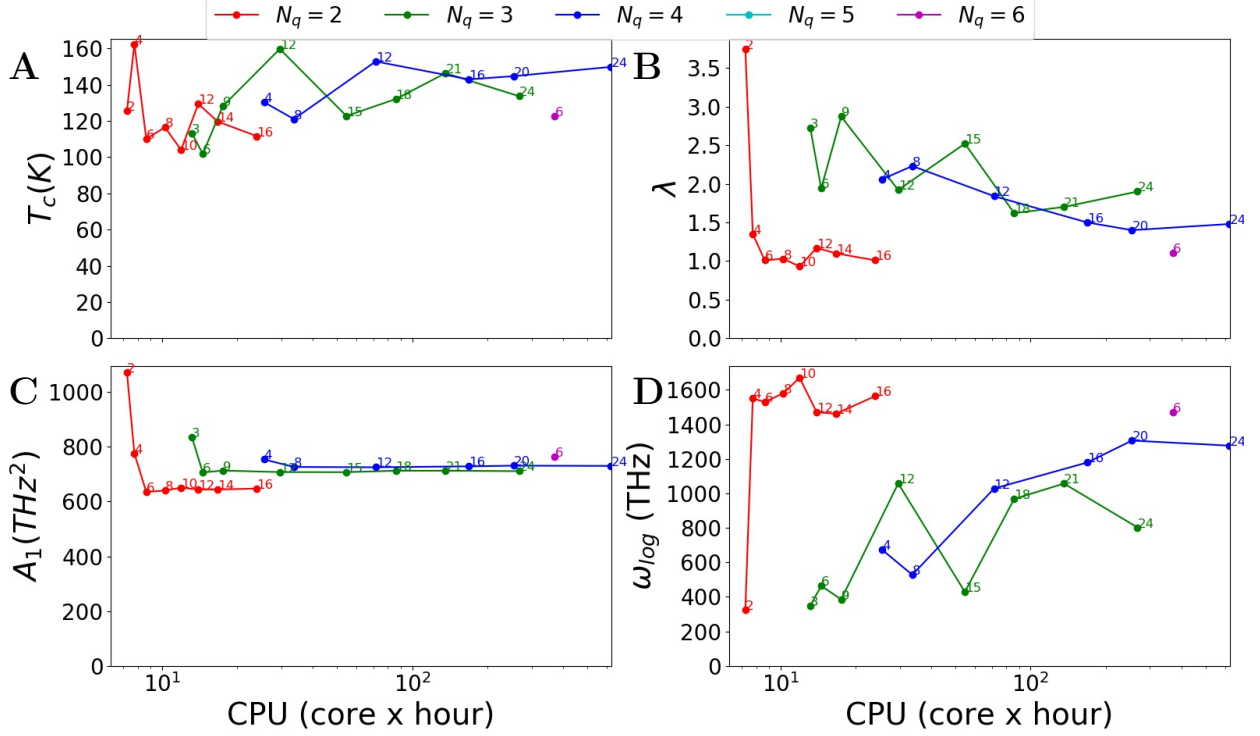
\begin{figure}[ht]
\begin{overpic}[width=\textwidth]{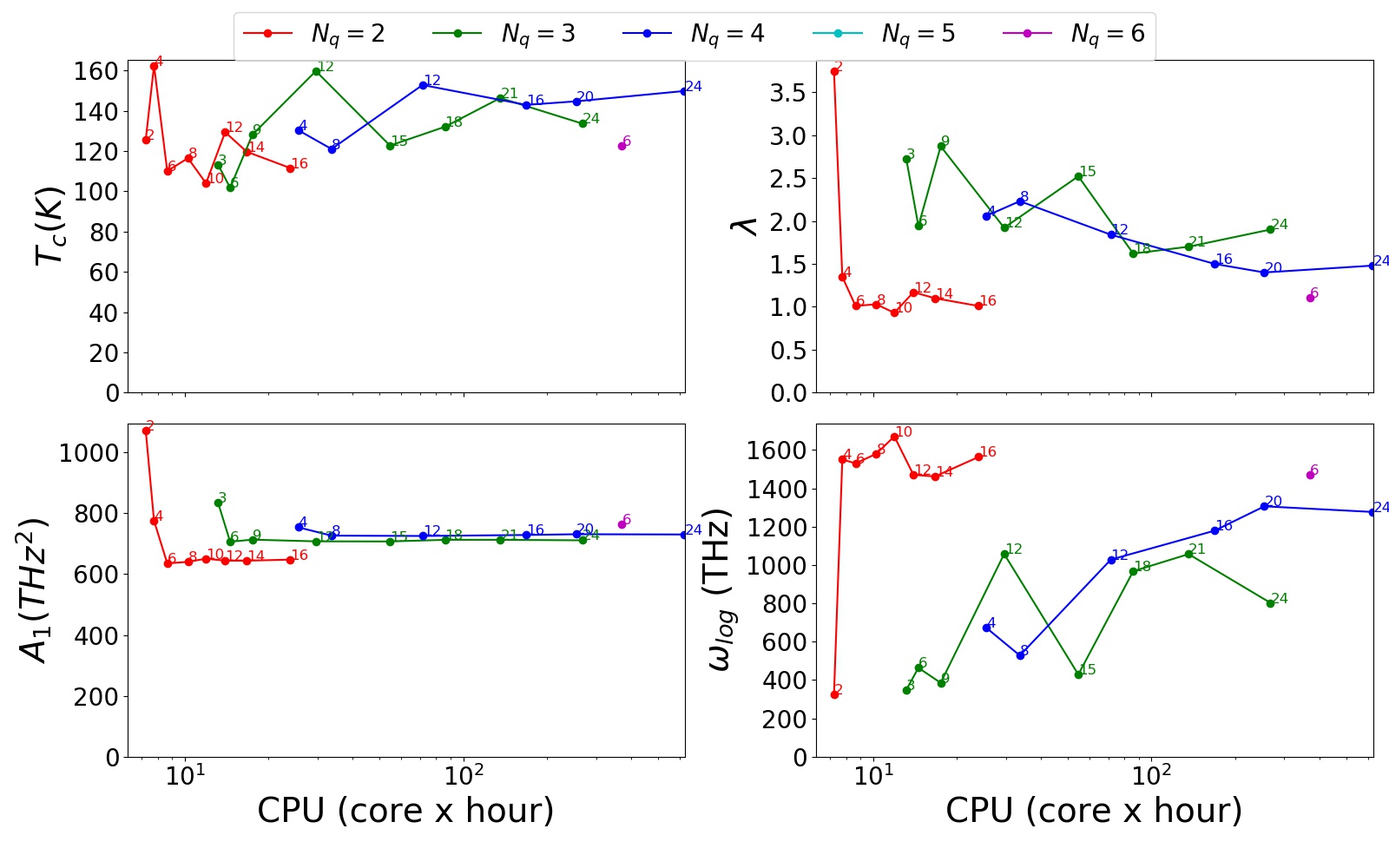}
\put (1,55) {\large\textbf A}
\put (51,55) {\large\textbf B}
\put (1,29) {\large\textbf C}
\put (51,29) {\large\textbf D}
\end{overpic}
\caption{$T_c$, $A_1$, $\lambda$, and $\omega_{\log}$ for $Fm\bar{3}m$ H$_3$S at 200~GPa as functions of computational cost. Red, green, blue, and cyan curves correspond to $N_q=2$, 3, 4, and 6, respectively. The labels at the data points give $N_k$. In this convergence test, $A_1$ approaches its converged value substantially faster than $T_c$, $\lambda$, and $\omega_{\log}$.}
\label{fig-A1-convergence}
\end{figure}

Figure~\ref{fig-A1-convergence} compares the convergence of $T_c$, $A_1$, $\lambda$, and $\omega_{\log}$ as functions of computational cost. In this test, $A_1$ reaches a near-converged value using approximately 8 core-hours, whereas a comparably converged $T_c$ requires approximately 200 core-hours. Thus, the convergence test provides a practical demonstration of the screening advantage of $A_1$ for this system. The exact saving depends on the material, pseudopotentials, smearing, and convergence criterion and should not be interpreted as a universal cost ratio. Additional convergence tests should be supplied in the Supplemental Material if a broader convergence claim is retained.

In the following sections, $A_1$ is used as an intermediate measure of electron--phonon perturbation strength to analyze periodic trends and prioritize $X_m$H$_n$ hydrides for subsequent dynamical-stability and EPC calculations.

\section*{Search for Cubic $X_m$H$_n$ Superconductors}

To establish a controlled dataset for examining trends across the periodic table, we selected 37 cubic $X_m$H$_n$ structural prototypes generated by the USPEX evolutionary algorithm. The prototypes satisfy $m\leq n$ and $m+n\leq 11$. For each prototype, we substituted $X$ with elements of atomic number $Z=2$--93 and performed structural, phonon, and EPC calculations at 200~GPa. This procedure produced more than 4,000 candidate structures, of which 133 are dynamically stable. The structural, electronic, vibrational, and superconducting properties of these 133 dynamically stable cubic systems are listed in Appendix~6.

Table~\ref{table-record} highlights the highest-$T_c$ dynamically stable cubic hydride identified in the present survey for each main group represented in the dataset and gives selected literature examples for context. No dynamically stable cubic $X_m$H$_n$ system was identified for Group~12 in the present survey at 200~GPa. The comparison is not intended as a comprehensive record of hydride superconductivity: the literature values were obtained for different structures, pressures, computational settings, and, in some cases, treatments of anharmonicity. Rather, it illustrates that the controlled cubic-prototype survey recovers several well-established high-$T_c$ hydrides while identifying additional candidates that merit higher-accuracy investigation. Before submission, the literature column should be updated against the current hydride literature and restricted to directly verifiable reference values.

\begin{table*}[t]
\caption{Representative high-$T_c$ hydrides for elements in different main groups. ``This work'' lists the highest-$T_c$ dynamically stable cubic $X_m$H$_n$ system identified in the present survey at 200~GPa. Selected literature values are provided only for qualitative context because the reported pressure, structure, computational methodology, and treatment of anharmonicity differ among studies.}
\label{table-record}
\centering
\begin{tabular}{@{\extracolsep{\fill}} c c c c c c c c}
\hline\hline
& \multicolumn{3}{c}{This work at 200~GPa} & \multicolumn{4}{c}{Selected literature example} \\
Group & Space & System & $T_c$ & Space & System & $T_c$  & Pressure \\
 & group & & (K) & group &  & (K) & (GPa) \\

\hline
1  & $Fm\bar{3}m$ & RbH         & 43.3  & $R\bar{3}m$ & LiH$_6$       & 82      & 300 \cite{xie2014superconductivity} \\
2  & $Im\bar{3}m$ & MgH$_6$     & 296.9 & $Im\bar{3}m$ & MgH$_6$       & 271     & 400 \cite{feng2015compressed} \\
3  & $Fm\bar{3}m$ & YH$_{10}$   & 294.0 & $Fm\bar{3}m$ & YH$_{10}$     & 305--326& 250 \cite{liu2017potential} \\
4  & $Im\bar{3}m$ & ZrH$_6$     & 129.6 & $P2_1/m$     & HfH$_2$       & 11--13  & 260 \cite{liu2015first} \\
5  & $Im\bar{3}m$ & NbH$_6$     & 174.9 & $Fdd2$       & TaH$_6$       & 124--136& 300 \cite{zhuang2017pressure} \\
6  & $Fm\bar{3}m$ & MoH          & 11.6  & $P6_3/mmc$   & CrH$_3$       & 37      & 81  \cite{yu2015pressure} \\
7  & $Pm\bar{3}m$ & TcH$_3$     & 44.6  & $I4/mmm$     & TcH$_2$       & 7--11   & 200 \cite{li2016crystal} \\
8  & $Fm\bar{3}m$ & FeH$_2$     & 6.8   & $I4/mmm$     & FeH$_5$       & 51      & 130 \cite{majumdar2017superconductivity} \\
9  & $Pm\bar{3}n$ & Ir$_2$H$_6$ & 65.6  & $Fm\bar{3}m$ & IrH           & 7       & 80  \cite{kim2011predicted} \\
10 & $Fm\bar{3}m$ & PtH          & 2.0   & $Fm\bar{3}m$ & PtH           & 25      & 77  \cite{errea2013first} \\
11 & $Fm\bar{3}m$ & AuH          & 56.7  & $Fm\bar{3}m$ & AuH           & 21      & 220 \cite{kim2011predicted} \\
13 & $F\bar{4}3m$ & AlH$_2$     & 105.0 & $Pbcn$       & B$_2$H$_6$    & 90--125 & 360 \cite{abe2011crystalline} \\
14 & $Pm\bar{3}n$ & Ge$_2$H$_6$ & 183.5 & $Pm\bar{3}n$ & Ge$_2$H$_6$   & 140     & 180 \cite{abe2011crystalline} \\
15 & $Fm\bar{3}m$ & PH           & 93.9  & $I4/mmm$     & PH$_2$        & 87      & 270 \cite{flores2016superconductivity} \\
16 & $Im\bar{3}m$ & SH$_3$      & 199.5 & $Im\bar{3}m$ & SH$_3$        & 191--204& 200 \cite{duan2014pressure} \\
17 & $Im\bar{3}m$ & BrH$_3$     & 165.9 & $R\bar{3}m$ & ClH$_2$       & 44--45  & 400 \cite{zeng2017emergence} \\
18 & $Fm\bar{3}m$ & KrH$_3$     & 91.6  & $Immm$       & XeH           & 29      & 100 \cite{yan2015structure} \\
\hline\hline
\end{tabular}
\end{table*}

\begin{figure}[ht]
\includegraphics[width=\textwidth]{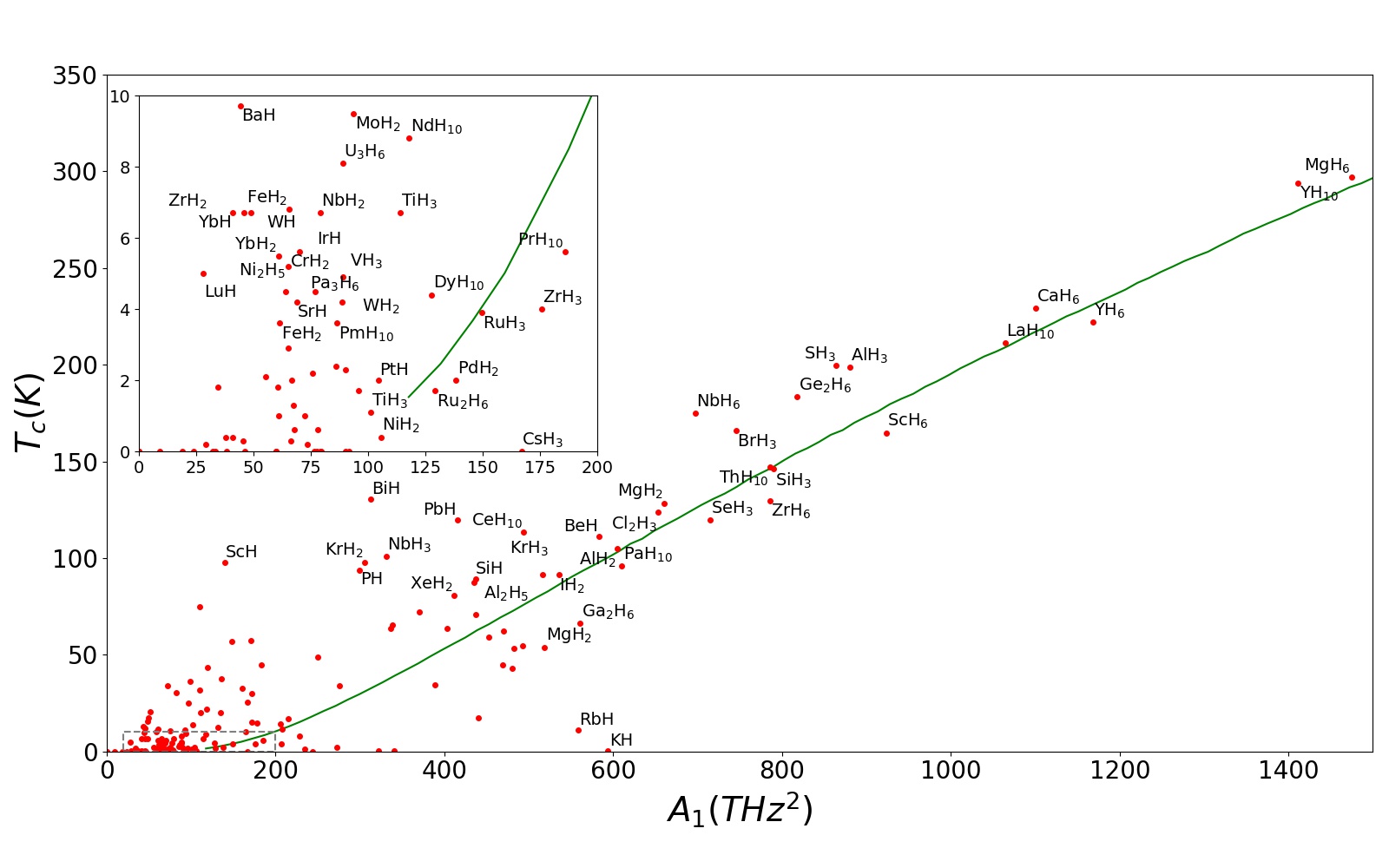}
\caption{Calculated $T_c$ and $A_1$ values for dynamically stable cubic $X_m$H$_n$ systems at 200~GPa. The inset enlarges the region of low $T_c$ and $A_1$. The green curve shows $\langle T_c(A_1)\rangle$, obtained from the random nonnegative $\alpha^2F(\omega)$ spectra described in the text.}
\label{fig-maxmin}
\end{figure}

Figure~\ref{fig-maxmin} compares the calculated $T_c$ and $A_1$ values for the dynamically stable cubic systems. The green curve gives the mean $T_c$ at a specified $A_1$, $\langle T_c(A_1)\rangle$, obtained from the random nonnegative $\alpha^2F(\omega)$ spectra described above. Across this controlled dataset, the Pearson correlation coefficient between the calculated $T_c$ and $A_1$ is 0.95. The correlation supports the use of $A_1$ as an intermediate screening descriptor for related cubic hydride frameworks; it does not imply that $A_1$ alone provides a quantitatively transferable predictor of $T_c$ across arbitrary structures, pressures, or phonon-frequency ranges.

\section*{Periodic Trends in Cubic $X_m$H$_n$ Hydrides}

A central question in the search for hydrogen-rich superconductors is how the electronic configuration of the nonhydrogen element $X$ influences the electron--phonon interaction and superconducting properties across the periodic table. Semenok \textit{et al.}\ suggested that high-$T_c$ hydrides are preferentially associated with elements having empty or nearly empty $d$ and $f$ shells, whereas increasing $d$- and $f$-electron occupation can suppress superconductivity \cite{semenok2018actinium}. Establishing such trends is difficult from literature data alone because reported compounds span different pressures, stoichiometries, structures, and computational protocols. Machine-learning and data-driven approaches have also been used to identify promising superconducting hydrides and to guide structure searches \cite{hutcheon2020predicting}. Such approaches are complementary to the present analysis, but their predictive performance depends on the coverage and consistency of the underlying training data. In particular, prior work examined the hydrogen-projected density-of-states ratio $N_{\mathrm{H}1s}(E_F)/N(E_F)$ as a low-cost ranking quantity and reported limited transferability across structurally distinct hydrides \cite{hutcheon2020predicting}. A controlled comparison within common structural prototypes is therefore useful for separating the effects of elemental substitution from those of the hydrogen framework and crystal structure.

Such a comparison is usually restricted by dynamical stability: only a small subset of elemental substitutions in a given prototype are dynamically stable at a fixed pressure. For example, at 200~GPa, the $Im\bar{3}m$ $X$H$_3$ prototype is dynamically stable only for selected elements, including S and Se, whereas the $Fm\bar{3}m$ $X$H$_{10}$ prototype is stable primarily for Y and selected lanthanide and actinide elements. The electron--perturbation coupling measure $A_1$ provides a common descriptor that can also be evaluated for structures with unstable harmonic phonons. We therefore calculate $A_1$ for a set of cubic $X_m$H$_n$ prototypes at 200~GPa with $X$ spanning atomic numbers $Z=2$--93.

\begin{figure}[ht]
\begin{overpic}[width=\textwidth]{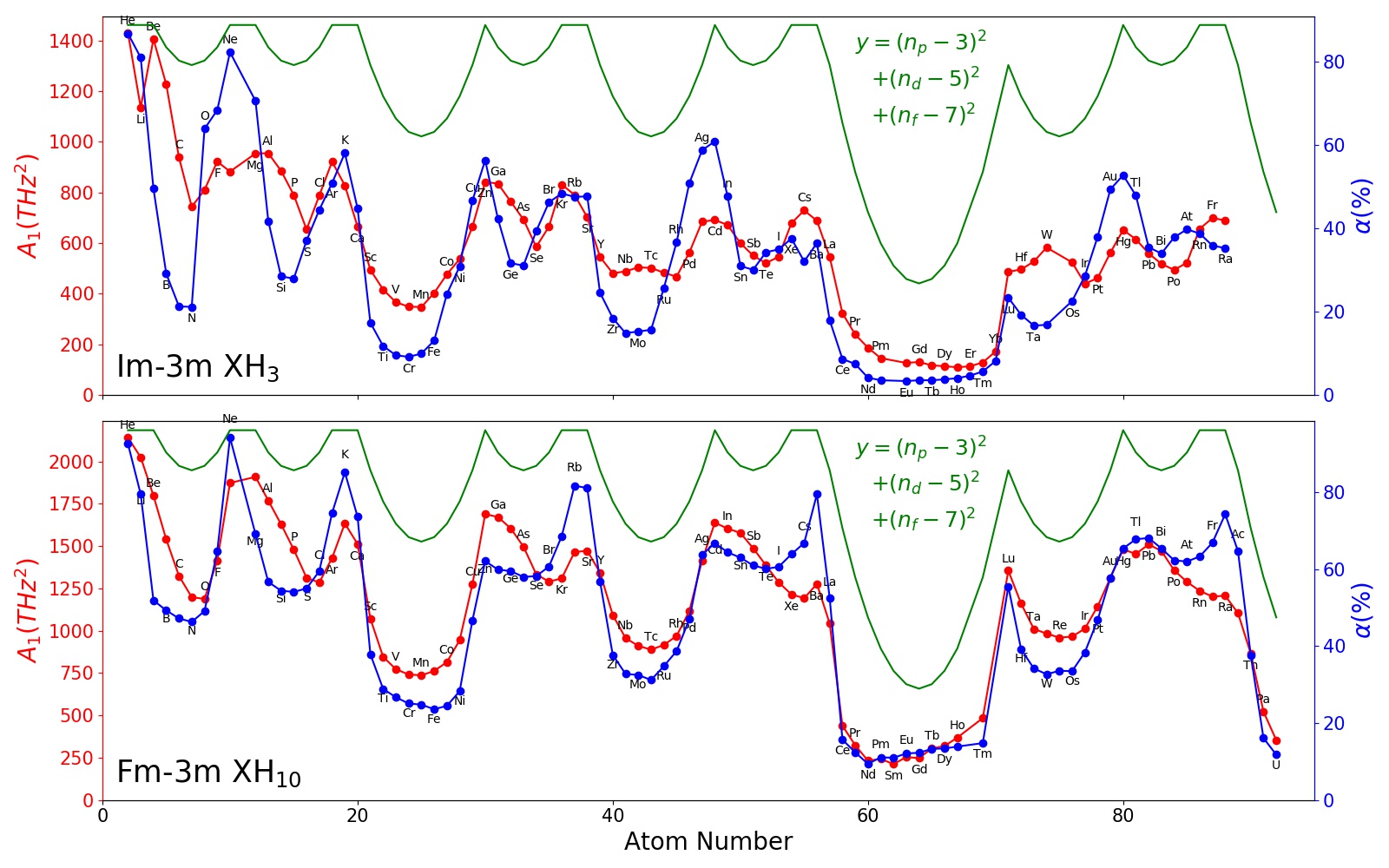}
\put (1,59) {\large\textbf A}
\put (1,29) {\large\textbf B}
\end{overpic}
\caption{Calculated $A_1$ and hydrogen contribution at the Fermi level, $\alpha$, for (A) $Im\bar{3}m$ $X$H$_3$ and (B) $Fm\bar{3}m$ $X$H$_{10}$ as functions of the atomic number $Z$ of $X$. Red and blue curves denote $A_1$ and $\alpha$, respectively. The green curve shows the empirical orbital-filling function $y(Z)=(n_p-3)^2+(n_d-5)^2+(n_f-7)^2$, included to guide the eye.}
\label{fig-A1_alpha}
\end{figure}

Figure~\ref{fig-A1_alpha} shows representative results for the $Im\bar{3}m$ $X$H$_3$ and $Fm\bar{3}m$ $X$H$_{10}$ prototypes. In both cases, $A_1(Z)$ displays pronounced periodic variations. Within a fixed prototype, local maxima tend to occur when the relevant $p$, $d$, or $f$ subshell is nearly empty or nearly filled, whereas local minima tend to occur near half filling. The overall variation is empirically described by
\begin{equation}
y(Z)=(n_p-3)^2+(n_d-5)^2+(n_f-7)^2,
\end{equation}
where $n_p$, $n_d$, and $n_f$ denote the nominal occupations of the $p$, $d$, and $f$ subshells of $X$, respectively. Similar periodic patterns are found for many of the other cubic prototypes considered here. This orbital-filling expression is used only as a compact description of the calculated trends; it is not intended as a universal predictor of superconductivity across unrelated structures or pressures.

The periodic behavior indicates that the valence electronic structure of $X$ strongly affects the electron--phonon perturbation associated with the hydrogen framework. To identify a directly calculable electronic descriptor, we define
\begin{equation}
\alpha=\frac{N_{\mathrm{H}1s}(E_F)}{N(E_F)},
\end{equation}
where $N(E_F)$ is the total electronic density of states at the Fermi energy and $N_{\mathrm{H}1s}(E_F)$ is its projection onto H-$1s$ orbitals. Thus, $\alpha$ measures the hydrogen character of the electronic states at the Fermi level. As shown in Fig.~\ref{fig-A1_alpha}, $A_1(Z)$ and $\alpha(Z)$ are strongly correlated for both representative prototypes. This result supports the physical picture that hydrogen-derived electronic states near $E_F$ are important for strong electron--phonon perturbations in hydrogen-rich compounds.

Although $\alpha$ provides a useful descriptor within a common structural prototype, its predictive ability is reduced when systems with different cubic hydrogen frameworks are combined. This limitation is consistent with the previously reported limited transferability of the hydrogen-projected density-of-states ratio across structurally distinct hydrides \cite{hutcheon2020predicting}. We therefore combine $\alpha$ with descriptors of Fermi-surface geometry and atomic density. For the subset of dynamically stable cubic systems for which the full descriptor analysis was performed, the Pearson correlation coefficient between $A_1$ and $\alpha$ is 0.746. We further consider the Fermi-surface descriptor $\beta$ and the atomic number density $\rho$. Here,
\begin{equation}
\beta=\frac{S_{\mathrm{FS}}}{S_{\mathrm{BZ}}}
\end{equation}
is the extremal Fermi-surface cross-sectional area expressed as a fraction of the corresponding cross-sectional area of the first Brillouin zone. For computational simplicity, the cross sections normal to the three reciprocal-lattice directions defined by the $b_1$--$b_2$, $b_1$--$b_3$, and $b_2$--$b_3$ planes are used. The atomic number density is
\begin{equation}
\rho=\frac{m+n}{V_{\mathrm{cell}}}.
\end{equation}
For this dataset, the correlation between $A_1$ and $\rho$ is 0.521, whereas the direct correlation between $A_1$ and $\beta$ is only 0.085. Despite the latter weak global correlation, $\beta$ is essential for identifying systems with appreciable hydrogen character at $E_F$ but a very small Fermi surface. For example, $Pm\bar{3}n$ Al$_2$H$_6$ has $\alpha=51.3\%$, comparable to that of LaH$_{10}$, but has a low calculated $T_c$ of 1.2~K and a very small $\beta$ of 8.1\%. For insulating systems, for which $\alpha$ is not defined at the Fermi level, assigning $\beta=0$ naturally excludes them from metallic-superconductor screening.

\begin{figure}[ht]
\begin{overpic}[width=\textwidth]{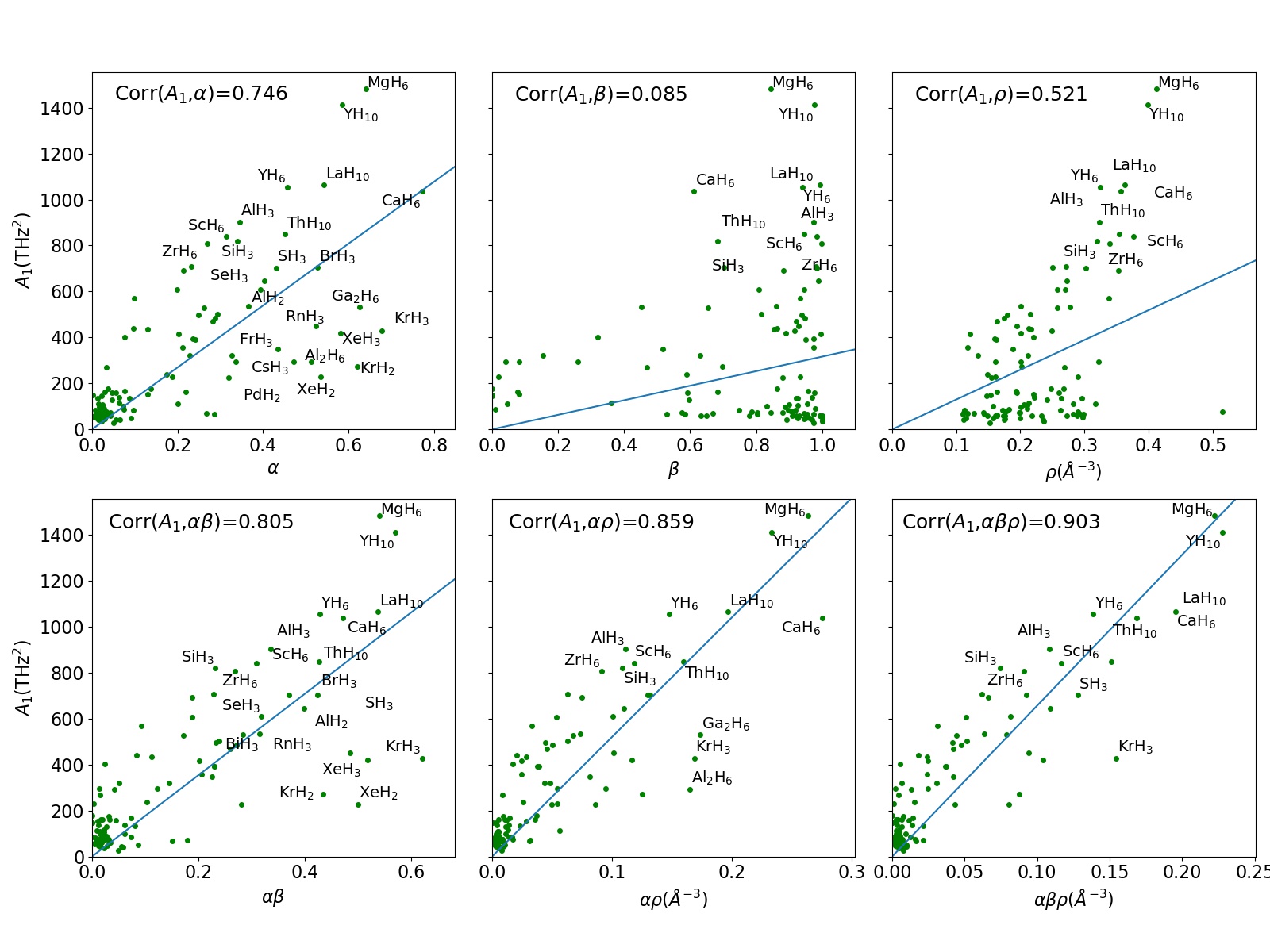}
\put (8,64) {\large\textbf A}
\put (40,64) {\large\textbf B}
\put (71,64) {\large\textbf C}
\put (8,30) {\large\textbf D}
\put (40,30) {\large\textbf E}
\put (71,30) {\large\textbf F}
\end{overpic}
\caption{Correlations of $A_1$ with (A) $\alpha$, (B) $\beta$, (C) $\rho$, (D) $\alpha\beta$, (E) $\alpha\rho$, and (F) $\alpha\beta\rho$ for the dynamically stable cubic systems included in the descriptor analysis. Blue lines are linear least-squares fits, and the corresponding Pearson correlation coefficients are shown in each panel.}
\label{fig-abrA1}
\end{figure}

Figure~\ref{fig-abrA1} compares $A_1$ with $\alpha$, $\beta$, $\rho$, and their products. Among the combinations considered, $\alpha\beta\rho$ gives the strongest correlation with $A_1$, with a Pearson coefficient of 0.903 for the present cubic dataset. This result motivates the empirical scaling
\begin{equation}
A_1 \approx h\alpha\beta\rho,
\end{equation}
where $h$ is a framework-dependent proportionality factor. The relation is used below as a physically motivated, low-cost screening criterion. It should not be interpreted as a quantitatively transferable prediction of $A_1$ or $T_c$ for arbitrary crystal structures, pressures, or phonon-frequency ranges.

\section*{Collision Model and Descriptor-Based Screening}

\begin{figure}[ht]
\begin{overpic}[width=\textwidth]{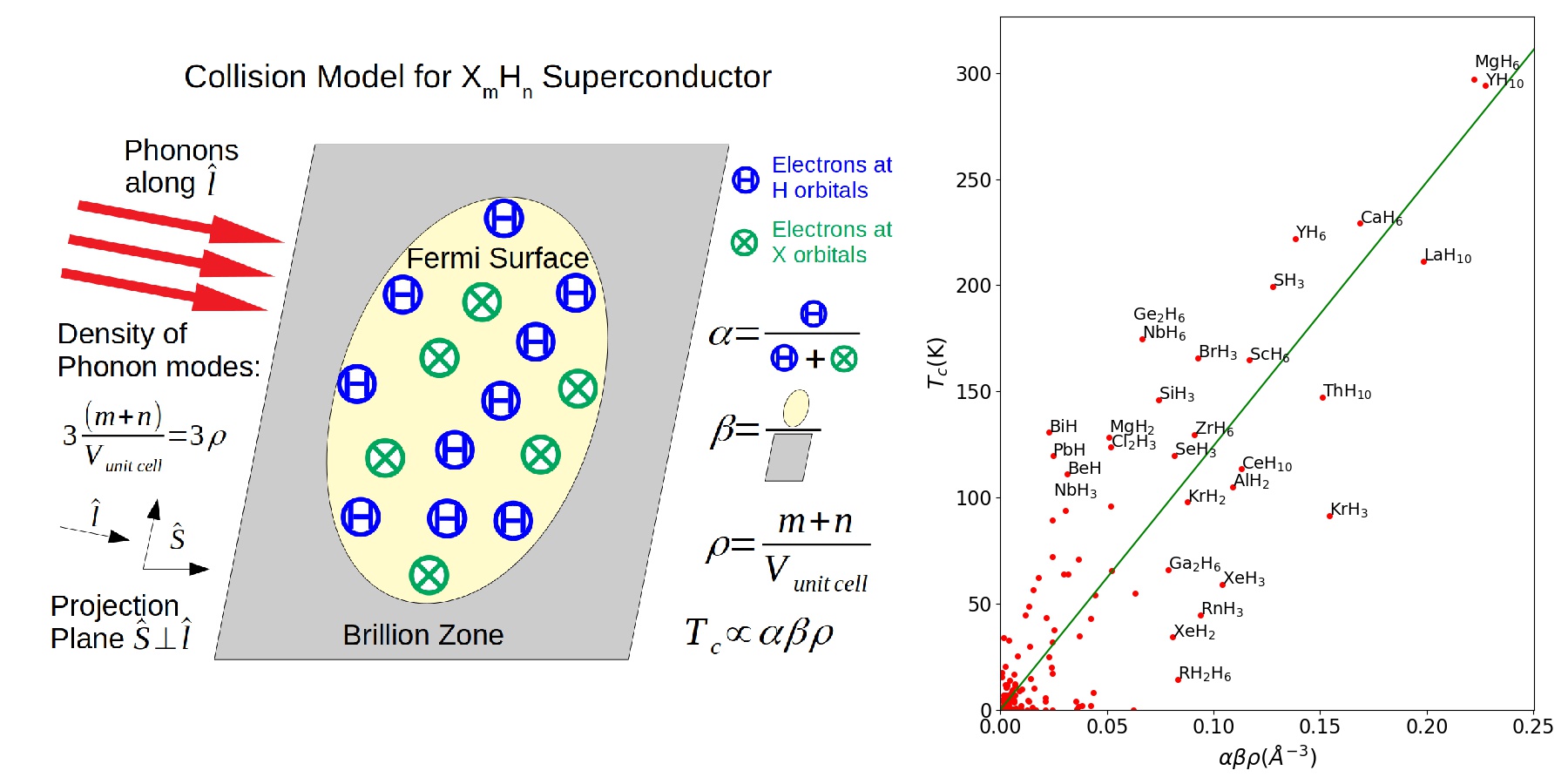}
\put (1,47) {\large\textbf A}
\put (59,47) {\large\textbf B}
\end{overpic}
\caption{(A) Schematic of the heuristic collision model. $\alpha$ is the H-$1s$ contribution to the electronic density of states at the Fermi level, $\beta$ measures the projected Fermi-surface area relative to the first Brillouin zone, and $\rho$ is the atomic number density. (B) Calculated $T_c$ as a function of $\alpha\beta\rho$ for the dynamically stable cubic systems included in the analysis at 200~GPa. The green line is a linear fit. The fitted relation is an empirical trend for this dataset, not a generally transferable $T_c$ prediction.}
\label{fig-collisionModel}
\end{figure}

The correlation analysis in the preceding section identifies the empirical scaling $A_1\approx h\alpha\beta\rho$ for the present cubic dataset. To rationalize this relation, we introduce a heuristic phase-space model. The model is intended to provide a physically transparent interpretation of the calculated trends; it is not a formal derivation of the electron--phonon interaction or a universally transferable model of $T_c$. A schematic representation is shown in Fig.~\ref{fig-collisionModel}(A).

In this picture, lattice perturbations are represented by localized phonon-like excitations propagating in reciprocal space. The number of phonon branches per unit volume scales as $3\rho$, where $\rho=(m+n)/V_{\mathrm{cell}}$ is the atomic number density. For a perturbation propagating along direction $\hat{\ell}$, the fraction of reciprocal-space trajectories that intersect the Fermi surface is approximated by
\begin{equation}
\beta_{\hat{\ell}}=\frac{\hat{S}_{\mathrm{FS}}}{\hat{S}_{\mathrm{BZ}}},
\end{equation}
where $\hat{S}_{\mathrm{FS}}$ and $\hat{S}_{\mathrm{BZ}}$ are the projected areas of the Fermi surface and first Brillouin zone, respectively, on a plane normal to $\hat{\ell}$. Averaging over propagation directions gives
\begin{equation}
\beta=\langle\beta_{\hat{\ell}}\rangle=\sum_{\hat{\ell}}p_{\hat{\ell}}\beta_{\hat{\ell}},
\end{equation}
where $p_{\hat{\ell}}$ is the directional weight. In the practical descriptor used here, $\beta$ is approximated from the cross sections normal to the three reciprocal-lattice directions, as defined above. Thus, $3\rho\beta$ provides a simple measure of the density of phonon branches with phase space for coupling to Fermi-level electronic states.

Let
\begin{equation}
\alpha_o=\frac{N_o(E_F)}{N(E_F)}
\end{equation}
be the fraction of the density of states at the Fermi level associated with orbital channel $o$. If $A_1^o$ denotes the corresponding channel-resolved electron--perturbation contribution, the model gives
\begin{equation}
A_1\propto 3\rho\beta\sum_o\alpha_o A_1^o.
\end{equation}
For the hydrogen-rich cubic systems considered here, the calculated trends indicate that the H-$1s$ channel makes a dominant contribution to this weighted sum. This approximation assumes that H-$1s$ states combine appreciable Fermi-level weight with a comparatively strong channel-resolved electron--perturbation contribution relative to the other orbital channels. Retaining only this dominant contribution yields
\begin{equation}
A_1\approx h\alpha\beta\rho,
\label{eq:A1abr}
\end{equation}
where $\alpha\equiv\alpha_{\mathrm{H}1s}$ is expressed as a fraction rather than a percentage and $h$ is an effective proportionality factor. Equation~\eqref{eq:A1abr} therefore recovers the empirical scaling identified above, but should be understood as a phase-space-motivated heuristic rather than an exact decomposition of the EPC matrix elements.

Figure~\ref{fig-collisionModel}(B) compares $T_c$ with $\alpha\beta\rho$ for the dynamically stable cubic systems used in the analysis at 200~GPa. Because these structures have broadly similar maximum phonon frequencies, with $\omega_{\max}=57\pm9$~THz, their calculated transition temperatures follow the approximate scaling
\begin{equation}
T_c\approx H\alpha\beta\rho,
\end{equation}
where $H=1275$~K\,\AA$^3$ for the present dataset. This fitted coefficient is framework- and pressure-dependent. It should not be regarded as a universal constant: it can change when the hydrogen framework, pressure, phonon-frequency scale, anharmonicity, or accuracy of the EPC calculation changes. In particular, this relation is useful principally for ranking related cubic systems, for which the hydrogen networks and vibrational scales are comparable.

The periodic trends can be understood qualitatively within this framework. Changing the valence configuration of $X$ changes the orbital composition of states near $E_F$ and therefore the H-$1s$ weight $\alpha$. Across a given structural prototype, the calculated maxima in $A_1$ often occur near empty or filled relevant $p$, $d$, or $f$ subshells, whereas minima frequently occur near half filling. This observation is consistent with changes in $X$--H hybridization, but it does not establish a universal one-to-one relation between orbital overlap, dynamical stability, and EPC. Structural stability also depends on bonding, enthalpy, pressure, and lattice dynamics. The practical design principle is therefore to seek compositions and structures that retain appreciable hydrogen-derived states at $E_F$ while remaining thermodynamically competitive and dynamically stable.

Within this heuristic framework, comparisons within a periodic group have similar limitations. In several main-group prototype series, larger $X$ atoms are associated with lower $\alpha$, $A_1$, and $T_c$, but these trends are not uniform across all groups or structures. In particular, transition-metal and $f$-electron systems involve more complex hybridization and cannot be described reliably by atomic size alone.

An important practical advantage is that $\alpha$, $\beta$, and $\rho$ can be obtained from relatively inexpensive self-consistent electronic-structure calculations, whereas reliable $A_1$ and $T_c$ calculations require phonon and EPC calculations. Thus, $\alpha\beta\rho$ provides a low-cost first-stage screening merit. For noncubic systems or structures with substantially different hydrogen frameworks, a small $\alpha\beta\rho$ reliably identifies low-priority candidates in the present tests, whereas a large value only identifies candidates that merit phonon and EPC calculations. It should not be interpreted as a quantitative prediction or a strict upper bound for $T_c$ outside the calibrated cubic dataset.

The descriptors also suggest possible routes for tuning candidate materials. For example, doping can shift $E_F$ and thereby alter $\alpha$ and the Fermi-surface geometry. If an electronic-structure calculation identifies a local maximum of $\alpha(E)$ at energy $E_o$, the rigid-band carrier concentrations required to shift the Fermi level to that energy are estimated as
\begin{equation}
n_e=\int_{E_F}^{E_o}N(E)\,dE \quad (E_o>E_F), \qquad
n_h=\int_{E_o}^{E_F}N(E)\,dE \quad (E_o<E_F).
\end{equation}
Such estimates are only a screening guide: explicit calculations are required to assess the effects of doping on structural stability, phonons, EPC, and the validity of the rigid-band approximation.

\section*{Three-Stage Search for High-$T_c$ Binary Hydride Superconductors at 50 and 200~GPa}

We applied a three-stage workflow to search binary hydrides across cubic and noncubic structural families at 50 and 200~GPa. The workflow uses progressively more expensive calculations: Stage~1 (S1) performs multiobjective evolutionary structure searches using enthalpic competitiveness and the low-cost electronic descriptor $\alpha\beta\rho$; Stage~2 (S2) evaluates harmonic phonons to assess dynamical stability; and Stage~3 (S3) calculates EPC and estimates $T_c$ using DFPT, as implemented in \textsc{Quantum ESPRESSO} \cite{giannozzi2009quantum}. Thus, full EPC calculations are reserved for the small subset of candidates that survive the preceding structural and dynamical-stability filters.

The direct objectives for a superconducting-hydride search would be dynamical stability and $T_c$. Both require phonon calculations, however, and $T_c$ further requires computationally demanding EPC calculations. We therefore use the enthalpy above the convex hull, $\Delta E=E-E_0$, and $\alpha\beta\rho$ as S1 objectives. Here, $E_0$ is the convex-hull enthalpy at the composition $n/(m+n)$, and $\beta$ is the mean projected Fermi-surface-area ratio evaluated on the three reciprocal-space planes spanned by $b_1$--$b_2$, $b_1$--$b_3$, and $b_2$--$b_3$. The two objectives represent a practical compromise between thermodynamic competitiveness and electronic phase space favorable for EPC. They are not substitutes for subsequent phonon and EPC calculations.

We used the Pareto-ranking implementation in USPEX to perform multiobjective evolutionary searches for $X_m$H$_n$ systems with $X=$ Li, Ne, Na, Mg, Al, Si, P, S, Cl, K, Ca, and Sc. After each structural relaxation, a Python post-processing script calculated $\alpha\beta\rho$ from the electronic structure and supplied this value to USPEX. The evaluation requires only a few milliseconds per relaxed structure and therefore adds negligible cost relative to the first-principles relaxation. Details of the USPEX implementation are provided in Appendix~7.

Structures satisfying $\Delta E<0.5$~eV/atom, $\alpha\beta\rho>0.1$~\AA$^{-3}$, Pareto rank $\mathrm{PR}<20$, and space-group number $\mathrm{SYMM}\geq2$ were advanced to S2. The symmetry criterion excludes fully asymmetric $P1$ candidates but is not intended as a stringent high-symmetry filter. The unit of the $\alpha\beta\rho$ threshold follows from expressing $\rho$ in \AA$^{-3}$ and treating $\alpha$ and $\beta$ as fractions. Harmonic phonons were calculated in S2 by the finite-displacement method using VASP \cite{kresse1996software} and Phonopy \cite{togo2015first}. Candidates without imaginary harmonic phonons were passed to S3. In S3, EPC calculations used a $\mathbf{k}$-point spacing of $0.05$~\AA$^{-1}$ and a $2\times2\times2$ $\mathbf{q}$-point mesh. The resulting $T_c$ values are intended as uniform, coarse-mesh screening estimates; final candidates require denser EPC sampling and, where relevant, consideration of anharmonic and other beyond-harmonic effects.

\begin{table}[t]
\caption{Numbers of structures evaluated in the three stages of the search workflow. S1 denotes the USPEX multiobjective search, S2 the harmonic-phonon stability filter, and S3 the coarse-mesh DFPT EPC calculations.}
\label{table-NumberCalcStage}
\centering
\begin{tabular}{c r r r r r r}
\hline\hline
& \multicolumn{3}{c}{200~GPa} & \multicolumn{3}{c}{50~GPa} \\
$X$ & $N_{\mathrm{S1}}$ & $N_{\mathrm{S2}}$ & $N_{\mathrm{S3}}$ & $N_{\mathrm{S1}}$ & $N_{\mathrm{S2}}$ & $N_{\mathrm{S3}}$ \\
\hline
Li & 3417 & 101 & 19 & 3340 & 68 & 2 \\
Ne & 3511 & 83 & 6 & 3602 & 26 & 0 \\
Na & 3668 & 85 & 12 & 3633 & 119 & 15 \\
Mg & 3649 & 96 & 25 & 3956 & 65 & 2 \\
Al & 3427 & 72 & 8 & 3919 & 88 & 1 \\
Si & 3219 & 72 & 20 & 3750 & 45 & 2 \\
P  & 4743 & 35 & 4 & 3463 & 73 & 3 \\
S  & 3683 & 75 & 10 & 3767 & 25 & 0 \\
Cl & 3620 & 40 & 4 & 3839 & 23 & 0 \\
K  & 3406 & 73 & 9 & 3456 & 46 & 5 \\
Ca & 3416 & 63 & 12 & 5094 & 6 & 0 \\
Sc & 3466 & 48 & 17 & 4135 & 44 & 9 \\
\hline
\textbf{Total} & \textbf{43225} & \textbf{843} & \textbf{146} & \textbf{45954} & \textbf{628} & \textbf{39} \\
\hline\hline
\end{tabular}
\end{table}

In total, 89,179 structures were considered in S1, 1,471 structures (1.65\%) proceeded to S2, and 185 structures (0.21\%) proceeded to S3 (Table~\ref{table-NumberCalcStage}). The high-$T_c$ candidates identified with the S3 screening calculations are listed in Tables~\ref{table-200GPa} and \ref{table-50GPa} for 200 and 50~GPa, respectively. Several of these candidates have not been examined previously and should be regarded as priorities for higher-accuracy structural, phonon, and EPC calculations, rather than as definitive predictions. Structures with the same chemical formula and closely matching crystal, electronic, and superconducting properties were treated as duplicates; only the highest-symmetry representative is listed. The complete candidate lists, including lower-$T_c$ systems and duplicates, are given in Appendix~8.

NaH$_6$ provides a useful validation example for the workflow. The S3 screen strongly prioritized several NaH$_6$ structures at 50 and 200~GPa, including cubic $Pm\bar{3}m$ and related lower-symmetry structures. Subsequent higher-accuracy calculations showed that $Pm\bar{3}m$ NaH$_6$ is dynamically stable down to 50~GPa and yields a harmonic Allen--Dynes estimate of $T_c=230$~K at 50~GPa, while the $R\bar{3}m$ and $C2/m$ phases yield lower transition temperatures at the same pressure. Structural relaxations further show that some initially identified low-symmetry NaH$_6$ structures transform to the $Pm\bar{3}m$ phase. A detailed analysis of the structural, vibrational, and superconducting properties of NaH$_6$ will be reported separately. This example illustrates the intended role of the present workflow: to efficiently identify promising candidates and structural families for subsequent high-accuracy investigation.

\begin{table*}[t]
\caption{Coarse S3 screening candidates at 200~GPa with $T_c\geq50$~K. Harmonic phonons, EPC, and $T_c$ were calculated using a $\mathbf{k}$-point spacing of $0.05$~\AA$^{-1}$ and a $2\times2\times2$ $\mathbf{q}$-point mesh. The $T_c$ values are screening estimates.}
\label{table-200GPa}
\centering
\begin{tabular}{@{\extracolsep{\fill}} c c c c c c c c c c}
\hline\hline
ID & Space & System & $T_c$ & $A_1$ & $\alpha$ & $\beta$ & $\rho$ & $N_F$(states/ & $\omega_{\max}$ \\
&Group &  & (K) & (THz$^2$) & (\%) & (\%) &($\text{\AA}^{-3}$) &Ry/spin/) & (THz)\\
& &  & & & & & &unit cell) &\\

\hline
Mg-3261 & $Im\bar{3}m$ & MgH$_6$ & 292 & 1729 & 75.3 & 88.4 & 0.419 & 2.830 & 75.1 \\
Na-2784 & $P2/m$ & NaH$_6$ & 272 & 1760 & 95.2 & 91.6 & 0.412 & 1.927 & 112.7 \\
Ca-281 & $Im\bar{3}m$ & CaH$_6$ & 237 & 1414 & 64.8 & 86.3 & 0.361 & 2.695 & 63.2 \\
K-1818 & $C2/m$ & KH$_{14}$ & 165 & 1525 & 90.5 & 87.1 & 0.455 & 3.207 & 117.0 \\
K-1974 & $Immm$ & KH$_{12}$ & 163 & 1029 & 81.8 & 89.8 & 0.432 & 2.625 & 121.5 \\
Li-2376 & $P\bar{1}$ & Li$_2$H$_6$ & 158 & 1119 & 77.8 & 80.0 & 0.447 & 1.689 & 104.2 \\
Al-1378 & $C2/m$ & AlH$_8$ & 151 & 1040 & 86.7 & 86.6 & 0.451 & 2.701 & 85.1 \\
K-2189 & $P\bar{1}$ & KH$_{12}$ & 143 & 1028 & 81.9 & 85.8 & 0.432 & 2.621 & 122.6 \\
S-2971 & $P\bar{1}$ & S$_2$H$_6$ & 133 & 696 & 41.7 & 96.4 & 0.309 & 5.325 & 64.9 \\
P-173 & $P\bar{1}$ & P$_2$H$_6$ & 127 & 700 & 48.8 & 92.5 & 0.314 & 5.545 & 88.3 \\
Mg-318 & $I4/mmm$ & MgH$_4$ & 113 & 1040 & 86.7 & 86.9 & 0.352 & 1.214 & 116.1 \\
S-2393 & $Im\bar{3}m$ & SH$_3$ & 108 & 475 & 42.7 & 85.0 & 0.305 & 2.680 & 52.5 \\
Li-3240 & $P\bar{1}$ & Li$_2$H$_{16}$ & 107 & 1004 & 87.8 & 93.2 & 0.534 & 3.389 & 109.1 \\
Ca-289 & $P\bar{1}$ & CaH$_8$ & 101 & 713 & 53.3 & 76.0 & 0.394 & 2.924 & 85.6 \\
Si-255 & $P\bar{1}$ & Si$_2$H$_6$ & 99 & 725 & 52.5 & 100.0 & 0.319 & 4.756 & 70.1 \\
Si-1899 & $Pm\bar{3}m$ & SiH$_3$ & 97 & 500 & 45.4 & 90.7 & 0.324 & 2.585 & 59.8 \\
Mg-475 & $P4/nmm$ & Mg$_2$H$_6$ & 96 & 614 & 87.1 & 94.9 & 0.318 & 3.641 & 64.2 \\
Na-1654 & $P4/mmm$ & NaH$_5$ & 87 & 909 & 97.5 & 98.0 & 0.391 & 1.294 & 117.3 \\
Ne-1790 & $I4$ & NeH$_8$ & 83 & 1027 & 93.7 & 41.8 & 0.460 & 0.347 & 146.5 \\
Cl-3280 & $Cm$ & ClH$_5$ & 77 & 1320 & 39.8 & 74.6 & 0.347 & 0.411 & 126.3 \\
Li-632 & $C2$ & LiH$_{10}$ & 77 & 930 & 93.4 & 91.4 & 0.529 & 1.734 & 111.8 \\
Mg-3048 & $P\bar{1}$ & MgH$_8$ & 76 & 886 & 78.4 & 83.3 & 0.443 & 2.197 & 87.1 \\
Na-1522 & $Cm$ & NaH$_9$ & 65 & 683 & 97.9 & 89.6 & 0.452 & 0.820 & 135.3 \\
Cl-1770 & $P6/mmm$ & ClH$_4$ & 61 & 656 & 35.8 & 100.0 & 0.329 & 3.243 & 106.2 \\
Mg-808 & $P\bar{1}$ & MgH$_{10}$ & 59 & 847 & 89.3 & 86.3 & 0.451 & 2.496 & 99.2 \\
Li-1319 & $P\bar{1}$ & LiH$_{10}$ & 57 & 905 & 92.7 & 87.2 & 0.528 & 1.693 & 122.3 \\
Mg-1086 & $P\bar{1}$ & Mg$_2$H$_{10}$ & 53 & 598 & 72.6 & 88.1 & 0.383 & 2.146 & 98.1 \\
\hline\hline
\end{tabular}
\end{table*}

\begin{table*}[t]
\caption{Coarse S3 screening candidates at 50~GPa with $T_c\geq50$~K. Harmonic phonons, EPC, and $T_c$ were calculated using a $\mathbf{k}$-point spacing of $0.05$~\AA$^{-1}$ and a $2\times2\times2$ $\mathbf{q}$-point mesh. The $T_c$ values are screening estimates.}
\label{table-50GPa}
\centering
\begin{tabular}{@{\extracolsep{\fill}} c c c c c c c c c c}
\hline\hline
ID & Space & System & $T_c$ & $A_1$ & $\alpha$ & $\beta$ & $\rho$ & $N_F$ (states/ & $\omega_{\max}$ \\
& Group & & (K) & (THz$^2$) & (\%) & (\%) & (\AA$^{-3}$) & Ry/spin/cell) & (THz) \\
\hline
Na-48 & $Pm\bar{3}m$ & NaH$_6$ & 236 & 1223 & 80.2 & 85.9 & 0.277 & 2.728 & 91.5 \\
K-2925 & $Immm$ & KH$_{12}$ & 149 & 746 & 86.7 & 77.9 & 0.278 & 3.262 & 112.8 \\
Na-752 & $R\bar{3}m$ & NaH$_6$ & 110 & 786 & 85.7 & 67.5 & 0.271 & 2.436 & 94.8 \\
Na-645 & $C2/m$ & NaH$_6$ & 96 & 797 & 85.6 & 91.7 & 0.271 & 2.411 & 96.5 \\
P-347 & $P\bar{1}$ & PH$_8$ & 76 & 389 & 73.7 & 85.5 & 0.278 & 3.297 & 122.8 \\
Sc-3044 & $C2/m$ & ScH$_{10}$ & 73 & 395 & 45.8 & 71.9 & 0.282 & 3.111 & 128.5 \\
Sc-805 & $C2/m$ & ScH$_6$ & 71 & 372 & 36.9 & 72.8 & 0.253 & 3.036 & 101.6 \\
Si-2744 & $C2/m$ & SiH$_8$ & 67 & 615 & 79.4 & 90.6 & 0.284 & 1.843 & 92.3 \\
Li-2877 & $P\bar{1}$ & LiH$_6$ & 50 & 665 & 78.4 & 71.7 & 0.333 & 1.756 & 97.5 \\
\hline\hline
\end{tabular}
\end{table*}

\clearpage

\section*{Conclusion}

In summary, we developed a physically motivated and computationally efficient framework for understanding and prioritizing candidate high-$T_c$ hydride superconductors. As an intermediate measure of electron--phonon perturbation strength, we introduced the spectral moment $A_1$, which can be evaluated for structures for which harmonic phonons are unstable and is less sensitive than $\lambda$, $\omega_{\log}$, and harmonic Allen--Dynes $T_c$ to the detailed phonon spectrum. It is therefore useful as a screening descriptor, but it does not establish superconductivity or physical realizability for a dynamically unstable structure. By examining more than 4,000 cubic $X_m$H$_n$ ($m\leq n$, $m+n\leq11$) structures generated using the USPEX evolutionary algorithm, we identify 133 dynamically stable systems and uncover pronounced periodic trends in their calculated electron--phonon and superconducting properties.

For compounds sharing a common structural prototype, $A_1$ exhibits systematic variations with the atomic number $Z$ of $X$: local maxima generally occur when the relevant $p$, $d$, or $f$ subshells are nearly empty or nearly filled, whereas local minima tend to occur near half filling. This behavior is described empirically by the orbital-filling function $(n_p-3)^2+(n_d-5)^2+(n_f-7)^2$, where $n_p$, $n_d$, and $n_f$ denote the nominal occupations of the corresponding subshells. Within the controlled cubic dataset, these trends show that the electronic character at the Fermi level, Fermi-surface geometry, atomic density, and hydrogen framework jointly influence the calculated electron--phonon perturbation strength.

To rationalize the empirical descriptor trends, we introduced a heuristic phase-space model in which $A_1\approx h\alpha\beta\rho$. Here, $\alpha$ is the H-$1s$ contribution to the density of states at the Fermi level, $\beta$ characterizes the projected Fermi-surface area, and $\rho$ is the atomic number density. The effective coefficient $h$ is framework- and pressure-dependent; it is not a universal material constant. For the related cubic hydrides considered here at 200~GPa, which have broadly comparable phonon-frequency scales, $\alpha\beta\rho$ provides a useful low-cost ranking metric. For structures with substantially different hydrogen frameworks, pressures, or phonon spectra, it should be used to prioritize explicit phonon and EPC calculations rather than to make a quantitative prediction of $T_c$.

These observations form the basis of the three-stage search strategy. Stage~1 uses $\alpha\beta\rho$ together with enthalpic competitiveness in a multiobjective evolutionary search; Stage~2 applies a harmonic-phonon stability filter; and Stage~3 performs coarse-mesh DFPT EPC calculations and harmonic Allen--Dynes $T_c$ screening for the surviving candidates. Applying this hierarchical workflow to more than 100,000 binary hydride structures at 50 and 200~GPa identifies several promising candidates beyond the cubic systems used to establish the descriptor relations. In particular, the workflow strongly prioritizes NaH$_6$ structural families at both pressures. Subsequent higher-accuracy calculations show that $Pm\bar{3}m$ NaH$_6$ is dynamically stable down to 50~GPa and has a harmonic Allen--Dynes estimate of $T_c\approx230$~K at that pressure. A detailed study of the structural, vibrational, and superconducting properties of NaH$_6$ will be reported separately.

More broadly, the combination of interpretable electronic descriptors, an intermediate electron--perturbation metric, and progressively more expensive stability and EPC calculations provides a scalable approach for exploring the large structural and compositional space of hydrogen-rich superconductors.

\section*{Acknowledgments}

T.C. acknowledges support from the NIST Fellowship Program during the initial stages of this work.

\bibliographystyle{unsrt}
\bibliography{main}

@article{errea2015high,
  title={High-pressure hydrogen sulfide from first principles: A strongly anharmonic phonon-mediated superconductor},
  author={Errea, Ion and Calandra, Matteo and Pickard, Chris J and Nelson, Joseph and Needs, Richard J and Li, Yinwei and Liu, Hanyu and Zhang, Yunwei and Ma, Yanming and Mauri, Francesco},
  journal={Physical Review Letters},
  volume={114},
  number={15},
  pages={157004},
  year={2015},
  publisher={APS}
}

@article{liu2017potential,
  title={Potential high-{$T_c$} superconducting lanthanum and yttrium hydrides at high pressure},
  author={Liu, Hanyu and Naumov, Ivan I and Hoffmann, Roald and Ashcroft, NW and Hemley, Russell J},
  journal={Proceedings of the National Academy of Sciences},
  volume={114},
  number={27},
  pages={6990--6995},
  year={2017},
  publisher={National Acad Sciences}
}

@article{somayazulu2019evidence,
  title={Evidence for superconductivity above 260~{K} in lanthanum superhydride at megabar pressures},
  author={Somayazulu, Maddury and Ahart, Muhtar and Mishra, Ajay K and Geballe, Zachary M and Baldini, Maria and Meng, Yue and Struzhkin, Viktor V and Hemley, Russell J},
  journal={Physical Review Letters},
  volume={122},
  number={2},
  pages={027001},
  year={2019},
  publisher={APS}
}

@article{allen1975transition,
  title={Transition temperature of strong-coupled superconductors reanalyzed},
  author={Allen, Ph B and Dynes, RC},
  journal={Physical Review B},
  volume={12},
  number={3},
  pages={905},
  year={1975},
  publisher={APS}
}

@incollection{allen1980dynamical,
  author    = {Allen, Philip B.},
  title     = {Neutron Spectroscopy of Superconductors},
  booktitle = {Dynamical Properties of Solids},
  editor    = {Horton, G. K. and Maradudin, A. A.},
  volume    = {3},
  pages     = {95},
  publisher = {North-Holland},
  address   = {Amsterdam},
  year      = {1980}
}

@article{togo2015first,
  title={First principles phonon calculations in materials science},
  author={Togo, Atsushi and Tanaka, Isao},
  journal={Scripta Materialia},
  volume={108},
  pages={1--5},
  year={2015},
  publisher={Elsevier}
}

@article{glass2006uspex,
  title={{USPEX}---Evolutionary crystal structure prediction},
  author={Glass, Colin W and Oganov, Artem R and Hansen, Nikolaus},
  journal={Computer Physics Communications},
  volume={175},
  number={11-12},
  pages={713--720},
  year={2006},
  publisher={Elsevier}
}

@article{semenok2018actinium,
  title={Actinium hydrides {AcH$_{10}$}, {AcH$_{12}$}, and {AcH$_{16}$} as high-temperature conventional superconductors},
  author={Semenok, Dmitrii V and Kvashnin, Alexander G and Kruglov, Ivan A and Oganov, Artem R},
  journal={The Journal of Physical Chemistry Letters},
  volume={9},
  number={8},
  pages={1920--1926},
  year={2018},
  publisher={ACS Publications}
}

@article{kresse1996software,
  author  = {Kresse, Georg and Furthm{\"u}ller, J{\"u}rgen},
  title   = {Efficient Iterative Schemes for \textit{ab initio} Total-Energy Calculations Using a Plane-Wave Basis Set},
  journal = {Physical Review B},
  volume  = {54},
  pages   = {11169--11186},
  year    = {1996},
  doi     = {10.1103/PhysRevB.54.11169}
}

@article{perdew1996generalized,
  title={Generalized gradient approximation made simple},
  author={Perdew, John P and Burke, Kieron and Ernzerhof, Matthias},
  journal={Physical Review Letters},
  volume={77},
  number={18},
  pages={3865},
  year={1996},
  publisher={APS}
}

@article{monkhorst1976special,
  title={Special points for Brillouin-zone integrations},
  author={Monkhorst, Hendrik J and Pack, James D},
  journal={Physical Review B},
  volume={13},
  number={12},
  pages={5188},
  year={1976},
  publisher={APS}
}

@article{giannozzi2009quantum,
  title={{QUANTUM ESPRESSO}: a modular and open-source software project for quantum simulations of materials},
  author={Giannozzi, Paolo and Baroni, Stefano and Bonini, Nicola and Calandra, Matteo and Car, Roberto and Cavazzoni, Carlo and Ceresoli, Davide and Chiarotti, Guido L and Cococcioni, Matteo and Dabo, Ismaila and others},
  journal={Journal of Physics: Condensed Matter},
  volume={21},
  number={39},
  pages={395502},
  year={2009},
  publisher={IOP Publishing}
}

@article{xie2014superconductivity,
  title={Superconductivity of lithium-doped hydrogen under high pressure},
  author={Xie, Yu and Li, Quan and Oganov, Artem R and Wang, Hui},
  journal={Acta Crystallographica Section C: Structural Chemistry},
  volume={70},
  number={2},
  pages={104--111},
  year={2014},
  publisher={International Union of Crystallography}
}

@article{feng2015compressed,
  title={Compressed sodalite-like {MgH$_6$} as a potential high-temperature superconductor},
  author={Feng, Xiaolei and Zhang, Jurong and Gao, Guoying and Liu, Hanyu and Wang, Hui},
  journal={RSC Advances},
  volume={5},
  number={73},
  pages={59292--59296},
  year={2015},
  publisher={Royal Society of Chemistry}
}

@article{liu2015first,
  title={First-principles study on the structural and electronic properties of metallic {HfH$_2$} under pressure},
  author={Liu, Yunxian and Huang, Xiaoli and Duan, Defang and Tian, Fubo and Liu, Hanyu and Li, Da and Zhao, Zhonglong and Sha, Xiaojing and Yu, Hongyu and Zhang, Huadi and others},
  journal={Scientific Reports},
  volume={5},
  pages={11381},
  year={2015},
  publisher={Nature Publishing Group}
}

@article{zhuang2017pressure,
  title={Pressure-stabilized superconductive ionic tantalum hydrides},
  author={Zhuang, Quan and Jin, Xilian and Cui, Tian and Ma, Yanbin and Lv, Qianqian and Li, Ying and Zhang, Huadi and Meng, Xing and Bao, Kuo},
  journal={Inorganic Chemistry},
  volume={56},
  number={7},
  pages={3901--3908},
  year={2017},
  publisher={ACS Publications}
}

@article{yu2015pressure,
  title={Pressure-driven formation and stabilization of superconductive chromium hydrides},
  author={Yu, Shuyin and Jia, Xiaojing and Frapper, Gilles and Li, Duan and Oganov, Artem R and Zeng, Qingfeng and Zhang, Litong},
  journal={Scientific Reports},
  volume={5},
  pages={17764},
  year={2015},
  publisher={Nature Publishing Group}
}

@article{li2016crystal,
  title={Crystal structures and superconductivity of technetium hydrides under pressure},
  author={Li, Xiaofeng and Liu, Hanyu and Peng, Feng},
  journal={Physical Chemistry Chemical Physics},
  volume={18},
  number={41},
  pages={28791--28796},
  year={2016},
  publisher={Royal Society of Chemistry}
}

@article{majumdar2017superconductivity,
  title={Superconductivity in {FeH$_5$}},
  author={Majumdar, Arnab and John, S Tse and Wu, Min and Yao, Yansun},
  journal={Physical Review B},
  volume={96},
  number={20},
  pages={201107},
  year={2017},
  publisher={APS}
}

@article{kim2011predicted,
  title={Predicted formation of superconducting platinum-hydride crystals under pressure in the presence of molecular hydrogen},
  author={Kim, Duck Young and Scheicher, Ralph H and Pickard, Chris J and Needs, RJ and Ahuja, Rajeev},
  journal={Physical Review Letters},
  volume={107},
  number={11},
  pages={117002},
  year={2011},
  publisher={APS}
}

@article{errea2013first,
  title={First-principles theory of anharmonicity and the inverse isotope effect in superconducting palladium-hydride compounds},
  author={Errea, Ion and Calandra, Matteo and Mauri, Francesco},
  journal={Physical Review Letters},
  volume={111},
  number={17},
  pages={177002},
  year={2013},
  publisher={APS}
}

@article{abe2011crystalline,
  title={Crystalline diborane at high pressures},
  author={Abe, Kazutaka and Ashcroft, NW},
  journal={Physical Review B},
  volume={84},
  number={10},
  pages={104118},
  year={2011},
  publisher={APS}
}

@article{flores2016superconductivity,
  title={Superconductivity in metastable phases of phosphorus-hydride compounds under high pressure},
  author={Flores-Livas, Jos{\'e} A and Amsler, Maximilian and Heil, Christoph and Sanna, Antonio and Boeri, Lilia and Profeta, Gianni and Wolverton, Chris and Goedecker, Stefan and Gross, EKU},
  journal={Physical Review B},
  volume={93},
  number={2},
  pages={020508},
  year={2016},
  publisher={APS}
}

@article{duan2014pressure,
  title={Pressure-induced metallization of dense {(H$_2$S)$_2$H$_2$} with high-{$T_c$} superconductivity},
  author={Duan, Defang and Liu, Yunxian and Tian, Fubo and Li, Da and Huang, Xiaoli and Zhao, Zhonglong and Yu, Hongyu and Liu, Bingbing and Tian, Wenjing and Cui, Tian},
  journal={Scientific Reports},
  volume={4},
  pages={6968},
  year={2014},
  publisher={Nature Publishing Group}
}

@article{zeng2017emergence,
  title={Emergence of novel hydrogen chlorides under high pressure},
  author={Zeng, Qingfeng and Yu, Shuyin and Li, Duan and Oganov, Artem R and Frapper, Gilles},
  journal={Physical Chemistry Chemical Physics},
  volume={19},
  number={12},
  pages={8236--8242},
  year={2017},
  publisher={Royal Society of Chemistry}
}

@article{yan2015structure,
  title={Structure, stability, and superconductivity of new {Xe--H} compounds under high pressure},
  author={Yan, Xiaozhen and Chen, Yangmei and Kuang, Xiaoyu and Xiang, Shikai},
  journal={The Journal of Chemical Physics},
  volume={143},
  number={12},
  pages={124310},
  year={2015},
  publisher={AIP Publishing}
}

@article{xie2022machine,
  author  = {Xie, S. R. and Quan, Y. and Hire, A. C. and Deng, B. and DeStefano, J. M. and Salinas, I. and Shah, U. S. and Fanfarillo, L. and Lim, J. and Kim, J. and Stewart, G. R. and Hamlin, J. J. and Hirschfeld, P. J. and Hennig, R. G.},
  title   = {Machine learning of superconducting critical temperature from {Eliashberg} theory},
  journal = {npj Computational Materials},
  volume  = {8},
  pages   = {14},
  year    = {2022},
  doi     = {10.1038/s41524-021-00666-7}
}

@article{wines2024machine,
  author  = {Wines, Daniel and Choudhary, Kamal},
  title   = {Data-driven design of high pressure hydride superconductors using {DFT} and deep learning},
  journal = {Materials Futures},
  volume  = {3},
  pages   = {025602},
  year    = {2024},
  doi     = {10.1088/2752-5724/ad4a94}
}

@article{pellegrini2024abinitio,
  author  = {Pellegrini, Camilla and Sanna, Antonio},
  title   = {Ab initio methods for superconductivity},
  journal = {Nature Reviews Physics},
  volume  = {6},
  pages   = {509--523},
  year    = {2024},
  doi     = {10.1038/s42254-024-00738-9}
}

@article{shipley2021high,
  author  = {Shipley, Alice M. and Hutcheon, Michael J. and Needs, Richard J. and Pickard, Chris J.},
  title   = {High-throughput discovery of high-temperature conventional superconductors},
  journal = {Physical Review B},
  volume  = {104},
  pages   = {054501},
  year    = {2021},
  doi     = {10.1103/PhysRevB.104.054501}
}

@article{saha2023high,
  author  = {Saha, Santanu and Di Cataldo, Simone and Giannessi, Federico and Cucciari, Alessio and von der Linden, Wolfgang and Boeri, Lilia},
  title   = {Mapping superconductivity in high-pressure hydrides: The {Superhydra} project},
  journal = {Physical Review Materials},
  volume  = {7},
  pages   = {054806},
  year    = {2023},
  doi     = {10.1103/PhysRevMaterials.7.054806}
}

@article{hutcheon2020predicting,
  author  = {Hutcheon, Michael J. and Shipley, Alice M. and Needs, Richard J.},
  title   = {Predicting novel superconducting hydrides using machine learning approaches},
  journal = {Physical Review B},
  volume  = {101},
  pages   = {144505},
  year    = {2020},
  doi     = {10.1103/PhysRevB.101.144505}
}

@article{quan2019compressed,
  author  = {Quan, Yundi and Ghosh, Soham S. and Pickett, Warren E.},
  title   = {How compressed hydrides produce room-temperature superconductivity},
  journal = {Physical Review B},
  volume  = {100},
  pages   = {184505},
  year    = {2019},
  doi     = {10.1103/PhysRevB.100.184505}
}

@incollection{allahyari2019multi,
  author    = {Allahyari, Zahed and Oganov, Artem R.},
  title     = {Multi-Objective Optimization as a Tool for Material Design},
  booktitle = {Handbook of Materials Modeling},
  editor    = {Andreoni, Wanda and Yip, Sidney},
  pages     = {1--15},
  publisher = {Springer},
  address   = {Cham},
  year      = {2019},
  doi       = {10.1007/978-3-319-50257-1_71-1}
}

\clearpage

\section*{Appendix 1. Computational Methods}

Cubic structural prototypes were generated using the USPEX evolutionary algorithm \cite{glass2006uspex}, interfaced with VASP. The search was restricted to the 36 cubic space groups (Nos.~195--230), unit cells containing at most 11 atoms, and hydrogen fractions $n/(m+n)\geq0.5$.

More than 4,000 candidate structures were structurally optimized at 200~GPa using density-functional theory as implemented in VASP. Core--valence interactions were treated with the projector-augmented-wave method, and exchange and correlation were described using the Perdew--Burke--Ernzerhof generalized-gradient approximation \cite{perdew1996generalized}. Brillouin-zone integrations used $8\times8\times8$ Monkhorst--Pack $\mathbf{k}$-point meshes \cite{monkhorst1976special}. Cubic symmetry was retained during these prototype relaxations. After relaxation, non-self-consistent calculations on $16\times16\times16$ $\mathbf{k}$-point meshes were used to obtain the projected density of states and Fermi-surface quantities required to evaluate $\alpha$, $\beta$, and $\rho$.

Phonon and EPC calculations were performed using DFPT, as implemented in \textsc{Quantum ESPRESSO}. Coarse calculations for the full prototype dataset used $8\times8\times8$ $\mathbf{k}$-point and $2\times2\times2$ $\mathbf{q}$-point meshes. Higher-accuracy calculations for the 133 harmonically stable cubic systems used $16\times16\times16$ $\mathbf{k}$-point and $4\times4\times4$ $\mathbf{q}$-point meshes. Gaussian broadenings of 0.10 and 0.04~Ry were used for the coarse and higher-accuracy calculations, respectively. The electron--perturbation measure $A_1$ was evaluated for all systems for which the required EPC quantities could be computed, whereas $T_c$ was evaluated from the Allen--Dynes formula only for dynamically stable systems. The reported $T_c$ values are harmonic Allen--Dynes estimates and use $\mu^*=0.1$.

The methodology and computational settings for the separate three-stage search are given in Appendix~7.

\section*{Appendix 2. Behavior of Spectral Moments in Dynamically Unstable Systems}

For an individual phonon mode, the contribution to the spectral moment $A_n$ can be written as
\begin{equation}
A_n(\mathbf{q},\nu)=\frac{1}{2\pi N_F}\gamma_{\mathbf{q}\nu}\omega_{\mathbf{q}\nu}^{n-1}.
\end{equation}
The negative moments $A_{-1}$ and $A_0$ contain inverse powers of the phonon frequency and can therefore become singular or numerically ill-conditioned when a mode softens to zero frequency. This behavior limits their use as screening quantities for structures with imaginary harmonic phonons. By contrast, $A_1(\mathbf{q},\nu)=\gamma_{\mathbf{q}\nu}/(2\pi N_F)$ contains no divergent inverse-frequency factor. Within the adopted linewidth-based construction, it can therefore remain finite as a harmonic mode softens. It does not, however, establish that an unstable structure is superconducting or physically realizable.

\begin{figure}[ht]
\centering
\includegraphics[width=\textwidth]{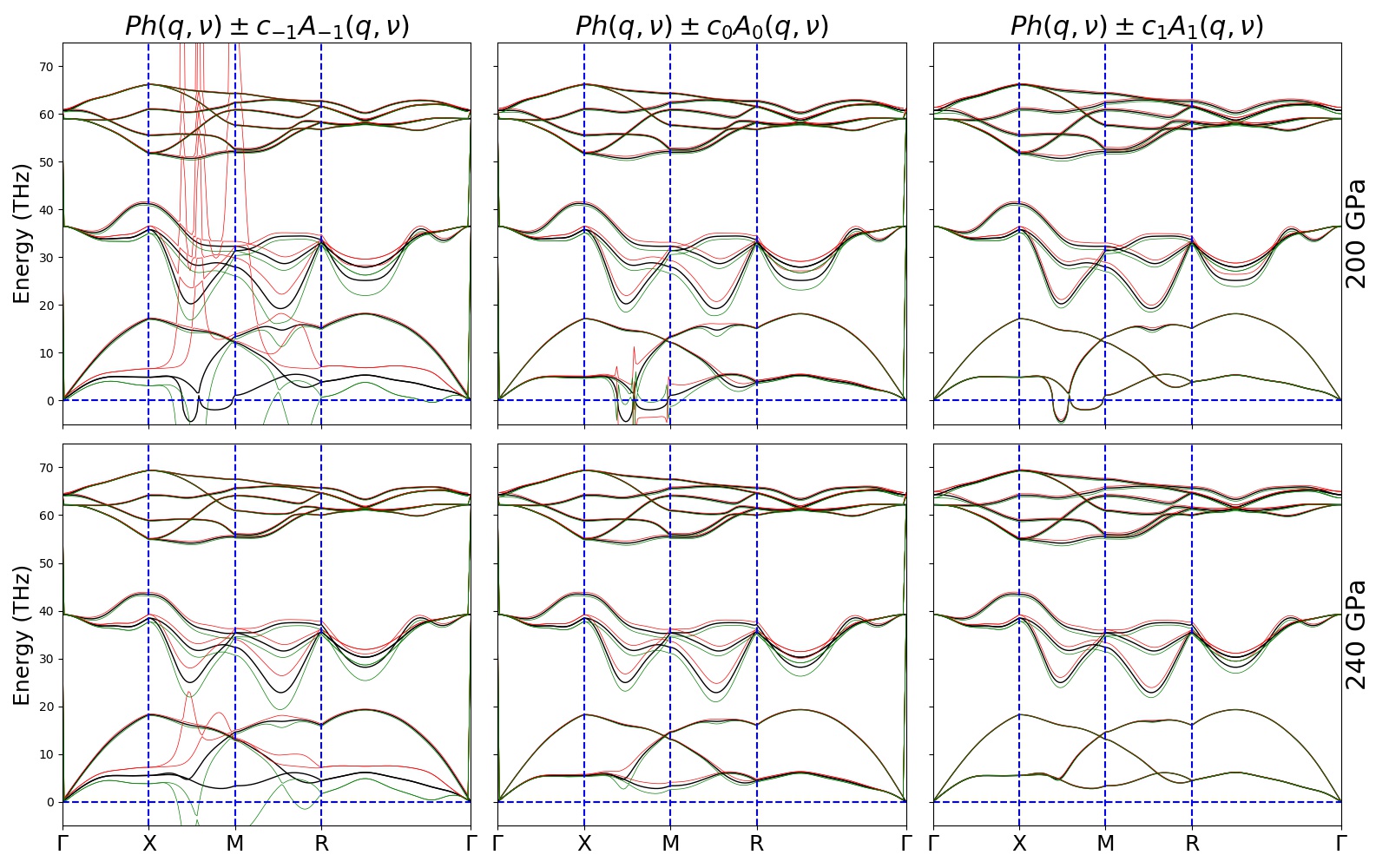}
\caption{Phonon dispersions and mode-resolved spectral-moment contributions $A_n(\mathbf{q},\nu)$ for $Fm\bar{3}m$ AlH$_3$ at 200 and 240~GPa. The phonon dispersions $\omega_{\mathbf{q}\nu}$ are shown as black solid curves. Red and blue curves show $\omega_{\mathbf{q}\nu}\pm c_nA_n(\mathbf{q},\nu)$, respectively, where $c_n$ are constant scaling factors selected for visualization. The 200-GPa calculation contains soft modes along $X$--$\Gamma$, whereas these modes have positive harmonic frequencies at 240~GPa.}
\label{fig-AlH3-An}
\end{figure}

Dynamically unstable structures may be stabilized by pressure, chemical substitution or doping, anharmonic effects, or relaxation to a lower-symmetry structure. As an illustrative example, cubic $Fm\bar{3}m$ AlH$_3$ has imaginary harmonic modes at 200~GPa but becomes harmonically stable at 240~GPa. Figure~\ref{fig-AlH3-An} compares its phonon dispersions and mode-resolved $A_n(\mathbf{q},\nu)$ values at the two pressures. At 200~GPa, soft modes occur along the $X$--$\Gamma$ path; at 240~GPa, these modes shift to small positive frequencies.

The $A_{-1}$ contribution grows strongly as the soft-mode frequency approaches zero and is consequently unsuitable for comparing the 200- and 240-GPa structures. The $A_0$ contribution is less singular but remains strongly enhanced near soft modes and is not well defined for imaginary harmonic frequencies. In contrast, $A_1$ remains finite within the adopted linewidth-based construction and varies smoothly across the soft-mode region. This example motivates the use of $A_1$ as an intermediate screening descriptor for electron--perturbation strength. Dynamical stability and quantitative superconducting properties must nevertheless be established through subsequent phonon and EPC calculations for the final candidates.

\section*{Appendix 3. $A_1$ Across the Calculated Crystal-Structure Prototypes}

\begin{figure}[ht]
\begin{overpic}[width=\textwidth]{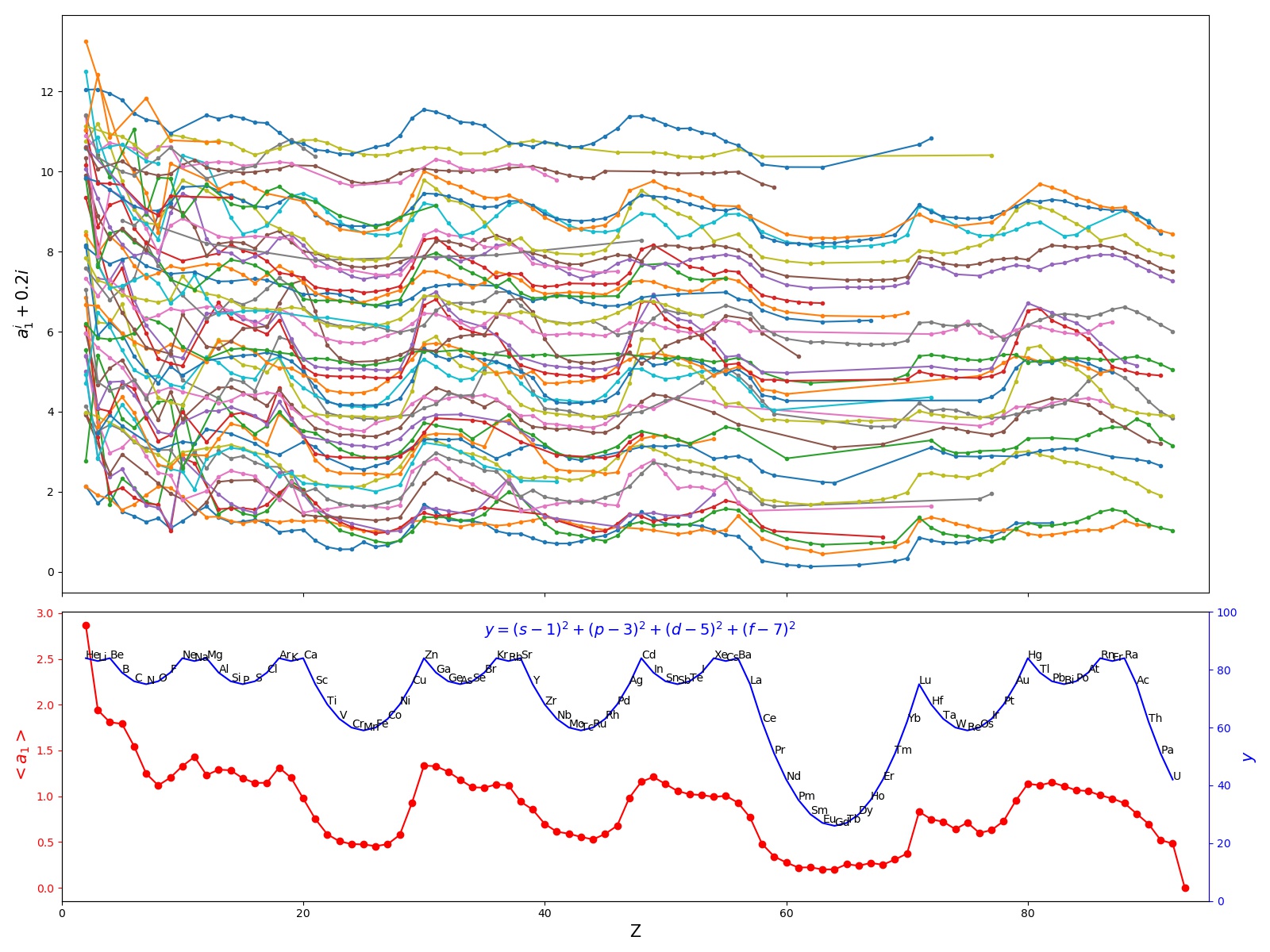}
\put (0,72) {\large\textbf A}
\put (0,25) {\large\textbf B}
\end{overpic}
\caption{(A) Normalized electron--perturbation measure $a_1^i(Z)$ for each of the 37 cubic structural prototypes. The curve for prototype $i$ is displaced by $0.2i$ for clarity. (B) Prototype-averaged normalized value $\langle a_1(Z)\rangle_i$ and the empirical orbital-filling function $y(Z)=(n_p-3)^2+(n_d-5)^2+(n_f-7)^2$.}
\label{fig-allA1}
\end{figure}

To examine periodic behavior across the full set of cubic prototypes, we calculated $A_1^i(Z)$ for each of the 37 prototypes $i$. For a given prototype, we define the normalized quantity
\begin{equation}
a_1^i(Z)=\frac{A_1^i(Z)}{\langle A_1^i(Z)\rangle_Z},
\end{equation}
where $\langle\cdots\rangle_Z$ denotes the average over the elemental substitutions for which $A_1$ could be evaluated. Figure~\ref{fig-allA1}(A) shows $a_1^i(Z)$ for all prototypes. For visual separation, the curve for prototype $i$ is shifted vertically by $0.2i$; these offsets have no physical meaning.

The prototype-averaged normalized trend,
\begin{equation}
\langle a_1(Z)\rangle_i=\frac{1}{N_i(Z)}\sum_i a_1^i(Z),
\end{equation}
where the sum includes the $N_i(Z)$ prototypes available at atomic number $Z$, is shown in Fig.~\ref{fig-allA1}(B). The averaged data exhibit a pronounced periodic variation. The function
\begin{equation}
y(Z)=(n_p-3)^2+(n_d-5)^2+(n_f-7)^2
\end{equation}
provides a compact empirical description of this trend, where $n_p$, $n_d$, and $n_f$ are the nominal $p$-, $d$-, and $f$-subshell occupations of $X$, respectively. This expression is included only as a guide to the calculated periodic behavior; it is not a universal predictor of EPC or superconductivity.

\begin{figure}[ht]
\centering
\includegraphics[width=\textwidth]{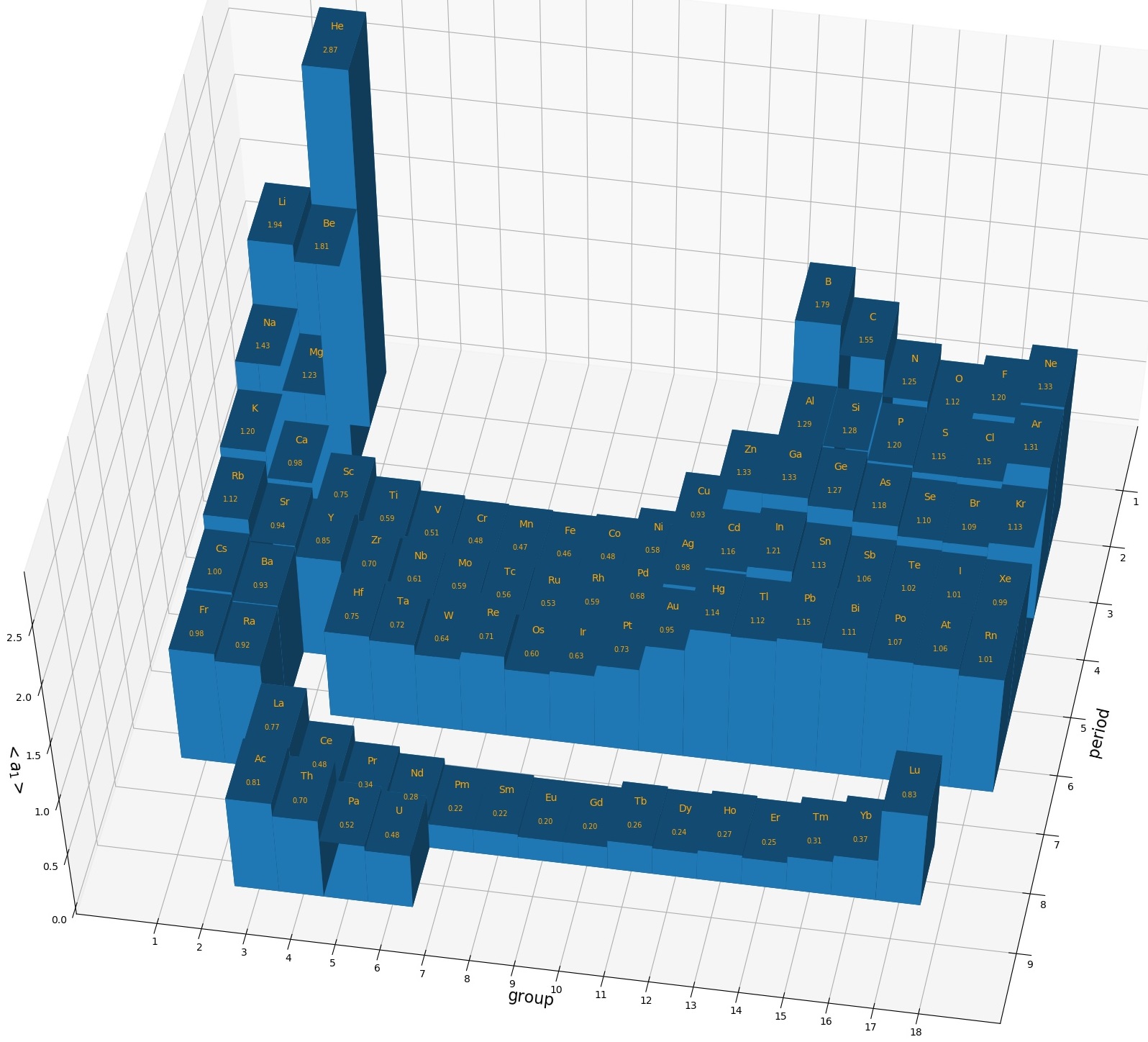}
\caption{Prototype-averaged normalized electron--perturbation measure $\langle a_1(Z)\rangle_i$ displayed on the periodic table.}
\label{fig-periodictable}
\end{figure}

Figure~\ref{fig-periodictable} replots $\langle a_1(Z)\rangle_i$ in periodic-table form.

\clearpage
\section*{Appendix 4. Illustrative Tuning by a Fermi-Level Shift}

\begin{figure}[ht]
\centering
\begin{overpic}[width=\textwidth]{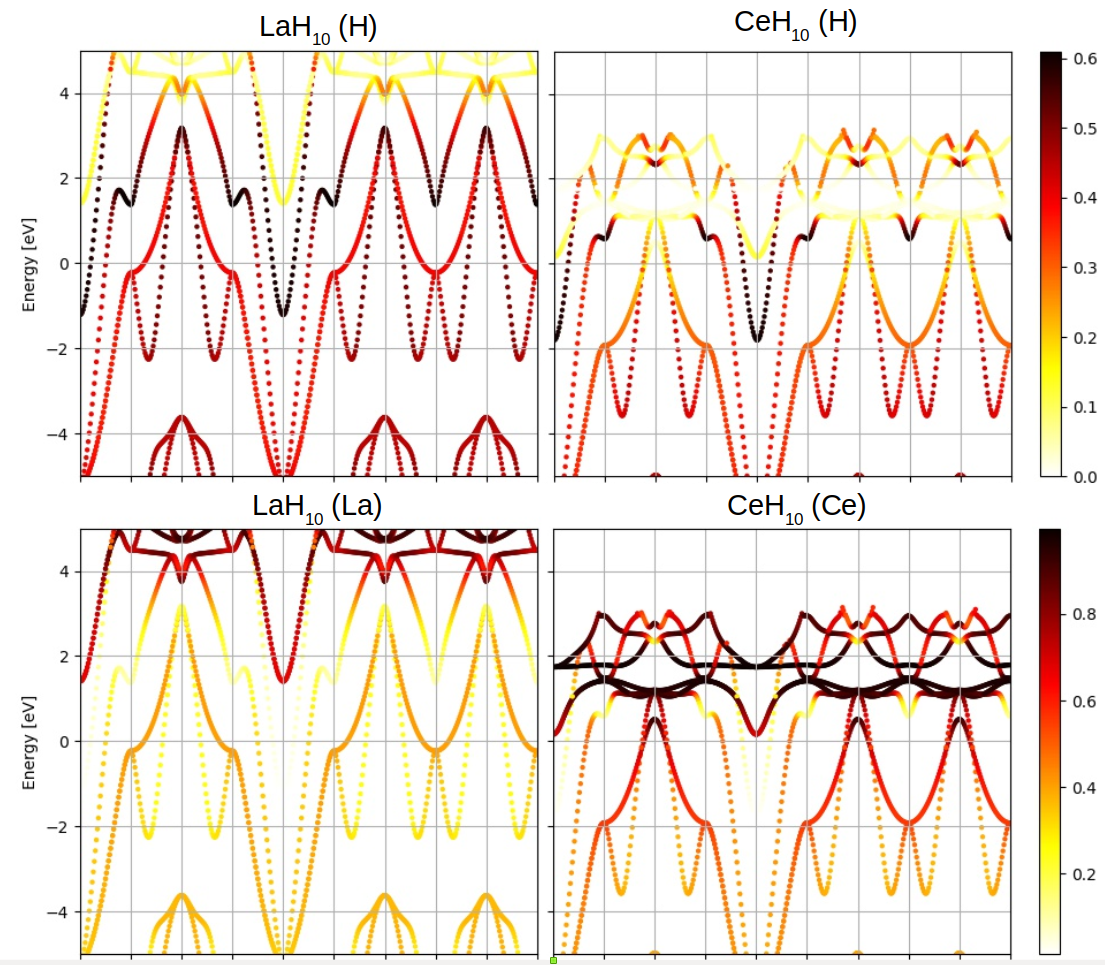}
\put (5,83) {\large\textbf A}
\put (50,83) {\large\textbf B}
\put (5,40) {\large\textbf C}
\put (50,40) {\large\textbf D}
\end{overpic}
\caption{Electronic band structures projected onto (A) H orbitals in LaH$_{10}$, (B) H orbitals in CeH$_{10}$, (C) La orbitals in LaH$_{10}$, and (D) Ce orbitals in CeH$_{10}$. The Fermi energy is set to zero.}
\label{fig-LaCe}
\end{figure}
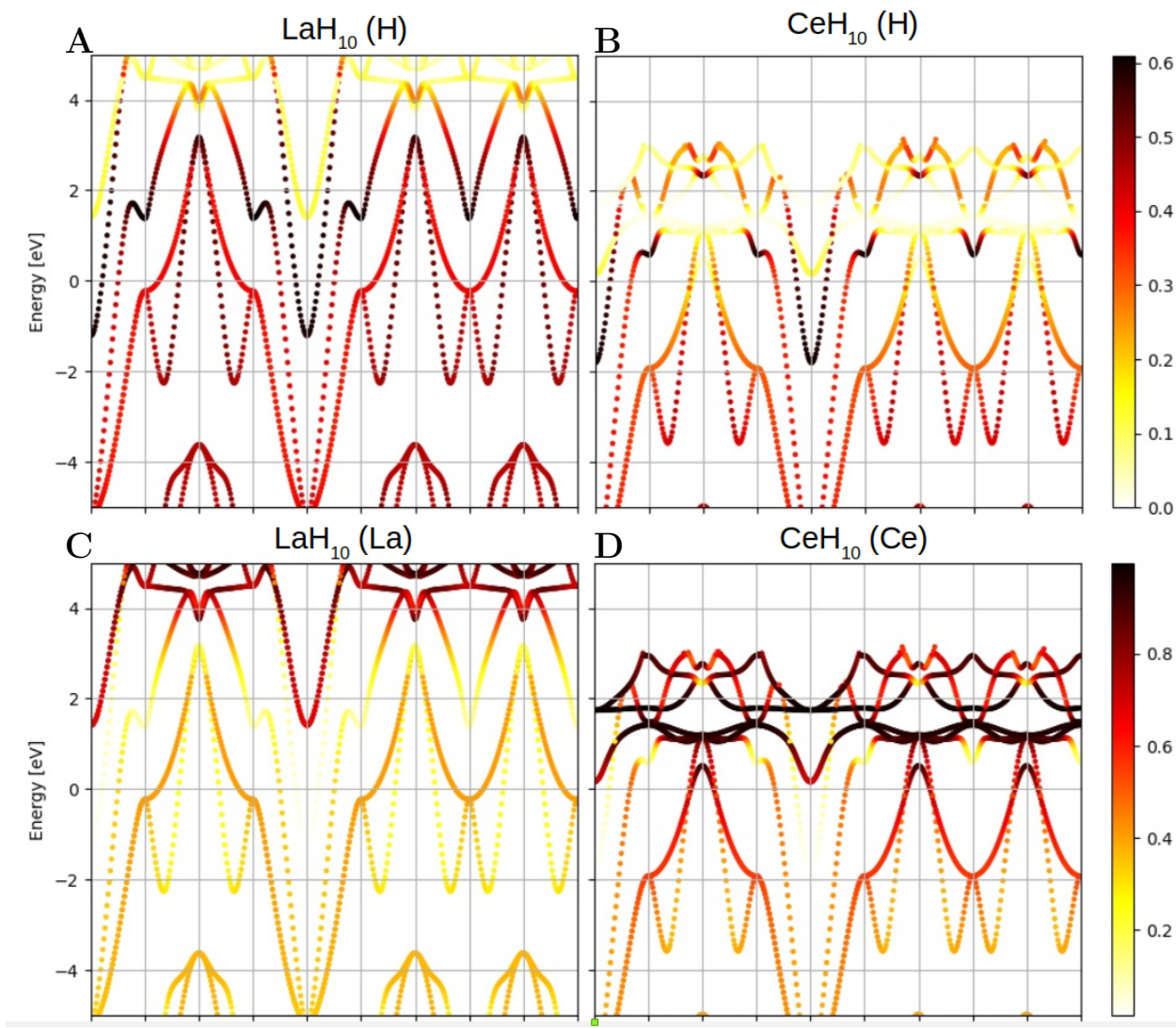

LaH$_{10}$ and CeH$_{10}$ provide an illustrative comparison because their electronic configurations differ principally through the occupation of a Ce-derived $4f$ state. In the present calculations, the superconducting transition temperature of CeH$_{10}$ is substantially lower than that of LaH$_{10}$. Figure~\ref{fig-LaCe} shows the H- and metal-projected electronic band structures. The Ce-derived $4f$ states modify the electronic structure near the Fermi level and substantially reduce the H-$1s$ contribution to the states at $E_F$ relative to LaH$_{10}$. This comparison is consistent with the descriptor analysis: within related hydrogen-rich structures, appreciable hydrogen-derived electronic character at $E_F$ is associated with stronger electron--phonon perturbations.

More generally, doping or chemical substitution can shift the Fermi level and alter both $\alpha$ and the Fermi-surface geometry. If $\alpha(E)$ has a local maximum at $E_o$, rigid-band estimates of the required carrier concentrations are
\begin{equation}
n_e=\int_{E_F}^{E_o}N(E)\,dE \quad (E_o>E_F), \qquad
n_h=\int_{E_o}^{E_F}N(E)\,dE \quad (E_o<E_F).
\end{equation}
These expressions are screening estimates only. Explicit calculations are required to determine whether doping is chemically feasible and to assess its effects on structural stability, lattice dynamics, EPC, and the validity of the rigid-band approximation.

\clearpage
\section*{Appendix 5. Scope and Limitations of $A_1\approx h\alpha\beta\rho$}

The descriptor relation $A_1\approx h\alpha\beta\rho$ is an empirical screening relation for the cubic dataset considered here. It is most useful for metallic structures with related hydrogen frameworks and broadly similar phonon-frequency scales. It should not be interpreted as an exact EPC identity or as a generally transferable $T_c$ prediction.

For an insulating electronic structure, $N(E_F)$ and the H-$1s$ fraction $\alpha=N_{\mathrm{H}1s}(E_F)/N(E_F)$ are not defined in the usual metallic sense. In the practical screening implementation, assigning $\beta=0$ excludes such systems from the metallic-superconductor candidate set. Near an insulator--metal transition, both the density of states at $E_F$ and the Fermi-surface geometry require particular care because small numerical changes can strongly affect the descriptors.

\begin{figure}[ht]
\centering
\begin{overpic}{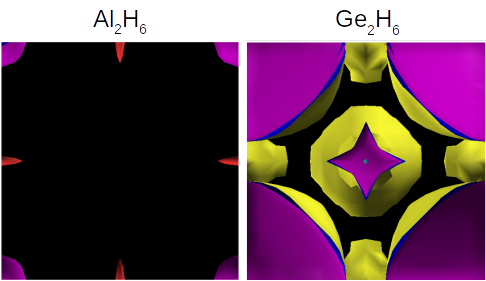}
\put (0,51) {\large\textbf A}
\put (50,51) {\large\textbf B}
\end{overpic}
\caption{Fermi surfaces of (A) $Pm\bar{3}n$ Al$_2$H$_6$ and (B) $Pm\bar{3}n$ Ge$_2$H$_6$.}
\label{fig-Pm-3n}
\end{figure}

Most dynamically stable cubic systems in the present 200-GPa dataset have relatively large projected Fermi-surface areas: 50\% have $\beta>90\%$, and 67\% have $\beta>80\%$. In exceptional cases, however, a small Fermi surface is the dominant factor limiting the electron--phonon phase space. For example, $Pm\bar{3}n$ Al$_2$H$_6$ has a larger H-$1s$ fraction at the Fermi level than $Pm\bar{3}n$ Ge$_2$H$_6$, but its calculated $A_1$ and $T_c$ are much smaller. Table~\ref{tab-AlGe} and Fig.~\ref{fig-Pm-3n} show that Al$_2$H$_6$ has a substantially smaller $\beta$ and $N_F$, consistent with its small Fermi surface.

\begin{table}[ht]
\caption{Calculated quantities for $Pm\bar{3}n$ Al$_2$H$_6$ and Ge$_2$H$_6$ at 200~GPa.}
\label{tab-AlGe}
\centering
\begin{tabular}{c c c c c c c c}
\hline\hline

System & $T_c$ & $A_1$ & $\alpha$& $\beta$& $\rho$ & $\omega_{\max}$  &$N_f$(states/ \\
& (K) & (THz$^2$) & (\%) & (\%) &($\text{\AA}^{-3}$) & (THz) &Ry/spin/cell) \\

\hline
Al$_2$H$_6$ & 1.2 & 235.2 & 51.3 & 8.1 & 0.332 & 67.3 & 0.983 \\
Ge$_2$H$_6$ & 183.5 & 818.2 & 23.2 & 98.2 & 0.271 & 54.8 & 7.626 \\
\hline\hline
\end{tabular}
\end{table}

As a descriptive sensitivity analysis, we fit the 133 dynamically stable cubic systems to
\begin{equation}
A_1=C\alpha^{k_1}\beta^{k_2}\rho^{k_3}\omega_{\max}^{k_4}N_F^{k_5}.
\end{equation}
The fitted exponents are $k_1=0.59\pm0.05$, $k_2=0.69\pm0.13$, $k_3=0.84\pm0.14$, $k_4=0.47\pm0.29$, and $k_5=-0.16\pm0.07$. Within this dataset, the fit supports the primary roles of $\alpha$, $\beta$, and $\rho$ and indicates a comparatively weak independent contribution from $N_F$ after these descriptors are included. The wide uncertainty in $k_4$ means that the apparent $\omega_{\max}$ dependence should not be overinterpreted.

A uniform rescaling of a fixed-shape Eliashberg spectrum illustrates why a phonon-frequency scale can matter. If $\alpha^2F(\omega)$ is replaced by $\alpha^2F(\omega/2)$, the characteristic frequencies and the harmonic Allen--Dynes $T_c$ scale upward by a factor of two, whereas $A_1$ scales upward by a factor of four. This mathematical rescaling does not establish a universal material relation between $T_c$, $A_1$, and $\omega_{\max}$, because real materials also change their EPC matrix elements, electronic structure, phonon spectrum, and stability under pressure or substitution. The modest spread $\omega_{\max}=57\pm9$~THz in the present cubic dataset helps explain why $\alpha\beta\rho$ provides a useful ranking descriptor there. Comparisons across different pressures or substantially different hydrogen networks require recalibration and direct phonon/EPC calculations.

\section*{Appendix 6. List of Dynamically Stable Cubic $X_m$H$_n$ Systems at 200~GPa}

The structural, electronic, vibrational, and superconducting quantities for the 133 dynamically stable cubic $X_m$H$_n$ systems at 200~GPa are listed below. The reported $T_c$ values are harmonic Allen--Dynes estimates using $\mu^*=0.1$.

\centering

\begin{longtable}{@{\extracolsep{\fill}} c c c c c c c c c @{}}
\hline

Space& System & $T_c$ & $A_1$ & $\alpha$& $\beta$& $\rho$&$N_f$(states/&$\omega_{max}$\\
Group &  & (K) & (THz$^2$) & (\%) & (\%) &($\text{\AA}^{-3}$) &Ry/spin/cell) & (THz)\\

\hline
\endhead
Im-3m	&	MgH$_6$	&	296.9	&	1,475.2	&	64.0	&	84.3	&	0.41	&	2.76	&	73.84	\\
Fm-3m	&	YH$_{10}$	&	294.0	&	1,411.4	&	58.4	&	97.7	&	0.40	&	4.94	&	69.99	\\
Im-3m	&	CaH$_6$	&	229.2	&	1,100.0	&	77.2	&	61.1	&	0.36	&	2.41	&	62.54	\\
Im-3m	&	YH$_6$	&	222.1	&	1,168.4	&	45.6	&	93.9	&	0.32	&	4.49	&	54.42	\\
Fm-3m	&	LaH$_{10}$	&	211.4	&	1,064.1	&	55.1	&	99.2	&	0.36	&	5.64	&	61.78	\\
Im-3m	&	SH$_3$	&	199.5	&	864.5	&	43.1	&	98.3	&	0.30	&	3.27	&	54.53	\\
Im-3m	&	NbH$_6$	&	174.9	&	697.1	&	21.3	&	88.3	&	0.35	&	6.39	&	63.01	\\
Im-3m	&	BrH$_3$	&	165.9	&	745.8	&	52.8	&	70.1	&	0.25	&	2.94	&	66.14	\\
Im-3m	&	ScH$_6$	&	164.8	&	923.3	&	31.4	&	98.4	&	0.38	&	5.34	&	64.50	\\
Fm-3m	&	ThH$_{10}$	&	147.3	&	786.0	&	45.1	&	94.6	&	0.35	&	3.94	&	64.51	\\
Pm-3m	&	SiH$_3$	&	146.0	&	790.4	&	34.0	&	68.2	&	0.32	&	2.59	&	61.69	\\
Fm-3m	&	BiH	&	130.7	&	312.8	&	20.1	&	95.2	&	0.12	&	3.25	&	38.54	\\
Im-3m	&	ZrH$_6$	&	129.6	&	785.7	&	26.9	&	99.7	&	0.34	&	6.47	&	61.71	\\
F-43m	&	MgH$_2$	&	128.4	&	660.7	&	19.9	&	94.6	&	0.27	&	3.09	&	68.14	\\
Pm-3m	&	Cl$_2$H$_3$	&	123.7	&	653.4	&	29.3	&	81.5	&	0.22	&	4.73	&	63.98	\\
Fm-3m	&	PbH	&	119.9	&	415.4	&	20.3	&	99.5	&	0.12	&	2.87	&	35.68	\\
Im-3m	&	SeH$_3$	&	119.7	&	715.0	&	39.3	&	80.9	&	0.26	&	3.17	&	65.85	\\
Fm-3m	&	CeH$_{10}$	&	113.5	&	493.7	&	30.1	&	98.4	&	0.38	&	5.40	&	66.49	\\
Fm-3m	&	BeH	&	111.1	&	583.4	&	9.9	&	93.2	&	0.34	&	1.41	&	52.83	\\
F-43m	&	AlH$_2$	&	105.0	&	604.7	&	40.3	&	98.9	&	0.27	&	1.82	&	65.73	\\
F-43m	&	KrH$_2$	&	97.9	&	305.4	&	62.1	&	69.8	&	0.20	&	2.97	&	49.69	\\
Fm-3m	&	PaH$_{10}$	&	95.9	&	610.0	&	15.8	&	88.2	&	0.37	&	10.45	&	63.69	\\
Fm-3m	&	PH	&	93.9	&	299.9	&	22.9	&	63.1	&	0.21	&	2.34	&	35.56	\\
Fm-3m	&	KrH$_3$	&	91.6	&	516.1	&	67.8	&	91.5	&	0.25	&	4.29	&	47.38	\\
Fm-3m	&	SiH	&	89.4	&	437.0	&	13.1	&	85.3	&	0.22	&	2.24	&	41.64	\\
Fm-3m	&	PoH	&	72.3	&	370.7	&	21.1	&	97.4	&	0.12	&	3.63	&	37.33	\\
F-43m	&	AtH$_2$	&	70.9	&	436.8	&	23.6	&	97.5	&	0.16	&	5.24	&	51.81	\\
Pm-3n	&	Ga$_2$H$_6$	&	66.1	&	560.8	&	62.5	&	45.3	&	0.28	&	5.15	&	61.77	\\
Pm-3n	&	Ir$_2$H$_6$	&	65.6	&	339.0	&	25.6	&	81.5	&	0.25	&	7.81	&	57.15	\\
Fm-3m	&	UH$_{10}$	&	63.8	&	403.7	&	8.3	&	100.0	&	0.38	&	21.85	&	65.02	\\
Fm-3m	&	RnH	&	63.8	&	336.1	&	29.2	&	90.3	&	0.11	&	5.09	&	42.87	\\
Fm-3m	&	MgH	&	62.4	&	470.2	&	9.7	&	86.3	&	0.21	&	2.80	&	52.06	\\
Fm-3m	&	XeH$_3$	&	59.0	&	453.1	&	58.1	&	89.0	&	0.20	&	3.88	&	46.49	\\
Fm-3m	&	AuH	&	56.7	&	148.6	&	17.5	&	59.0	&	0.15	&	1.97	&	32.71	\\
Fm-3m	&	BiH$_3$	&	54.7	&	492.3	&	36.6	&	86.2	&	0.20	&	2.88	&	53.30	\\
Fm-3m	&	MgH$_2$	&	53.9	&	518.9	&	26.2	&	65.5	&	0.26	&	1.62	&	70.79	\\
Fm-3m	&	ZrH$_3$	&	48.8	&	249.8	&	6.1	&	93.6	&	0.24	&	5.15	&	60.38	\\
Fm-3m	&	RnH$_3$	&	44.8	&	469.6	&	52.3	&	92.7	&	0.19	&	3.47	&	50.07	\\
Pm-3m	&	TcH$_3$	&	44.6	&	183.4	&	4.6	&	97.6	&	0.26	&	7.59	&	58.16	\\
Fm-3m	&	RbH	&	43.3	&	119.8	&	28.1	&	54.1	&	0.14	&	2.31	&	58.34	\\
F-43m	&	BiH$_2$	&	43.1	&	480.8	&	28.2	&	92.2	&	0.16	&	3.05	&	57.20	\\
Pm-3n	&	Os$_2$H$_6$	&	37.6	&	135.7	&	10.3	&	95.4	&	0.25	&	5.87	&	58.09	\\
F-43m	&	PoH$_2$	&	34.6	&	389.4	&	24.2	&	94.9	&	0.16	&	3.59	&	54.41	\\
F-43m	&	XeH$_2$	&	34.2	&	275.9	&	53.5	&	93.3	&	0.16	&	3.66	&	49.11	\\
Fm-3m	&	NbH	&	33.9	&	72.1	&	1.1	&	90.1	&	0.16	&	6.16	&	52.92	\\
F-43m	&	ZrH$_2$	&	32.6	&	160.3	&	2.9	&	68.3	&	0.19	&	5.52	&	59.79	\\
Pm-3m	&	Ce$_2$H$_8$	&	31.8	&	109.9	&	5.2	&	90.8	&	0.51	&	18.30	&	60.43	\\
F-43m	&	ScH$_2$	&	29.8	&	171.7	&	6.3	&	96.8	&	0.22	&	8.07	&	61.43	\\
Pm-3m	&	NbH$_3$	&	25.3	&	166.8	&	3.7	&	86.4	&	0.25	&	5.88	&	67.13	\\
P-43m	&	Pd$_2$H$_5$	&	25.0	&	97.2	&	16.5	&	99.6	&	0.14	&	8.37	&	68.73	\\
Fm-3m	&	VH	&	20.5	&	51.8	&	1.4	&	78.9	&	0.20	&	7.23	&	56.15	\\
Fm-3m	&	RhH$_2$	&	20.2	&	111.8	&	14.0	&	78.6	&	0.22	&	3.38	&	59.96	\\
P-43m	&	Fe$_2$H$_5$	&	17.4	&	49.4	&	0.4	&	100.0	&	0.19	&	42.23	&	78.82	\\
Pm-3n	&	Zr$_2$H$_6$	&	17.2	&	440.1	&	11.5	&	89.6	&	0.24	&	8.53	&	61.42	\\
F-43m	&	HfH$_2$	&	16.8	&	215.2	&	5.7	&	59.1	&	0.20	&	4.39	&	61.52	\\
P-43m	&	Co$_2$H$_5$	&	15.6	&	48.8	&	0.4	&	100.0	&	0.19	&	41.98	&	75.99	\\
Pm-3m	&	MoH$_3$	&	14.8	&	178.0	&	5.8	&	97.9	&	0.25	&	9.65	&	59.37	\\
Pm-3n	&	Rh$_2$H$_6$	&	14.4	&	205.4	&	40.6	&	76.3	&	0.27	&	5.04	&	58.22	\\
F-43m	&	NbH$_2$	&	14.0	&	101.9	&	2.4	&	89.9	&	0.21	&	4.65	&	61.29	\\
F-43m	&	TiH$_2$	&	12.3	&	132.3	&	4.6	&	59.7	&	0.24	&	5.82	&	63.02	\\
Fm-3m	&	TaH	&	11.8	&	45.8	&	1.8	&	78.8	&	0.15	&	5.16	&	50.34	\\
Fm-3m	&	MoH	&	11.6	&	61.0	&	1.9	&	94.9	&	0.16	&	5.01	&	49.55	\\
F-43m	&	TaH$_2$	&	11.2	&	92.8	&	1.7	&	88.8	&	0.20	&	4.46	&	61.79	\\
Pm-3m	&	RbH	&	11.2	&	558.7	&	32.6	&	15.5	&	0.13	&	0.99	&	50.95	\\
Pm-3m	&	CaH	&	10.5	&	75.5	&	1.6	&	90.5	&	0.18	&	4.49	&	53.45	\\
Fm-3m	&	CrH	&	10.4	&	58.6	&	1.3	&	93.1	&	0.22	&	6.49	&	55.07	\\
Pm-3m	&	OsH$_3$	&	10.3	&	165.1	&	7.0	&	90.9	&	0.25	&	4.49	&	56.96	\\
Fm-3m	&	BaH	&	9.7	&	44.3	&	9.2	&	94.9	&	0.12	&	4.49	&	55.53	\\
Fm-3m	&	MoH$_2$	&	9.5	&	93.7	&	2.8	&	97.8	&	0.21	&	4.83	&	63.38	\\
Fm-3m	&	NdH$_{10}$	&	8.8	&	117.9	&	2.2	&	100.0	&	0.40	&	36.67	&	67.77	\\
Fm-3m	&	U$_3$H$_6$	&	8.1	&	88.9	&	2.8	&	100.0	&	0.18	&	32.42	&	60.71	\\
F-43m	&	RnH$_2$	&	7.9	&	228.3	&	31.9	&	87.9	&	0.16	&	3.28	&	53.23	\\
Fm-3m	&	FeH$_2$	&	6.8	&	65.4	&	0.5	&	100.0	&	0.29	&	12.17	&	63.87	\\
Fm-3m	&	WH	&	6.7	&	49.0	&	1.7	&	92.6	&	0.16	&	4.61	&	46.09	\\
Fm-3m	&	TiH$_3$	&	6.7	&	114.0	&	1.3	&	92.0	&	0.30	&	6.00	&	62.82	\\
Fm-3m	&	NbH$_2$	&	6.7	&	79.1	&	1.1	&	88.0	&	0.20	&	3.99	&	64.41	\\
Fm-3m	&	YbH	&	6.7	&	45.7	&	4.7	&	97.5	&	0.15	&	7.15	&	56.34	\\
Fm-3m	&	PrH$_{10}$	&	5.6	&	185.8	&	5.4	&	99.0	&	0.40	&	22.01	&	68.13	\\
Fm-3m	&	IrH	&	5.6	&	70.2	&	3.4	&	99.4	&	0.16	&	6.64	&	36.37	\\
F-43m	&	YbH$_2$	&	5.5	&	60.9	&	4.1	&	57.5	&	0.20	&	4.18	&	61.63	\\
Fm-3m	&	CrH$_2$	&	5.2	&	65.1	&	1.1	&	99.2	&	0.27	&	7.44	&	65.45	\\
Fm-3m	&	LuH	&	5.0	&	28.0	&	1.1	&	63.2	&	0.15	&	2.63	&	55.43	\\
Fm-3m	&	VH$_3$	&	4.9	&	89.0	&	1.6	&	95.5	&	0.32	&	5.77	&	62.23	\\
P-43m	&	Ni$_2$H$_5$	&	4.5	&	64.2	&	6.8	&	100.0	&	0.19	&	12.16	&	75.90	\\
Fm-3m	&	Pa$_3$H$_6$	&	4.5	&	77.0	&	3.6	&	100.0	&	0.17	&	22.52	&	60.03	\\
F-43m	&	WH$_2$	&	4.2	&	88.7	&	2.5	&	92.6	&	0.21	&	3.35	&	60.88	\\
Pm-3m	&	SrH	&	4.2	&	68.8	&	2.6	&	84.4	&	0.14	&	3.73	&	50.01	\\
Pm-3m	&	ZrH$_3$	&	4.0	&	175.7	&	7.4	&	77.8	&	0.23	&	4.15	&	67.31	\\
Pm-3m	&	RuH$_3$	&	3.9	&	149.5	&	8.7	&	92.5	&	0.26	&	5.18	&	59.75	\\
Pm-3n	&	Y$_2$H$_6$	&	3.8	&	206.5	&	36.5	&	43.8	&	0.22	&	2.09	&	61.51	\\
Fm-3m	&	PmH$_{10}$	&	3.6	&	86.4	&	1.6	&	100.0	&	0.40	&	45.43	&	68.51	\\
F-43m	&	FeH$_2$	&	3.6	&	61.5	&	0.5	&	100.0	&	0.30	&	11.63	&	61.18	\\
Fm-3m	&	LaH	&	2.3	&	90.1	&	3.3	&	94.5	&	0.13	&	7.05	&	42.97	\\
F-43m	&	MnH$_2$	&	2.2	&	75.8	&	1.1	&	97.4	&	0.29	&	8.35	&	63.50	\\
Fm-3m	&	TcH	&	2.1	&	55.3	&	1.0	&	92.0	&	0.17	&	4.13	&	47.49	\\
Fm-3m	&	FrH$_3$	&	2.1	&	272.8	&	43.5	&	51.8	&	0.19	&	1.69	&	57.57	\\
Fm-3m	&	PdH$_2$	&	2.0	&	138.4	&	44.5	&	40.4	&	0.21	&	0.87	&	58.18	\\
F-43m	&	CrH$_2$	&	2.0	&	66.7	&	1.0	&	95.5	&	0.28	&	6.19	&	64.41	\\
Fm-3m	&	PtH	&	2.0	&	104.5	&	7.3	&	83.3	&	0.16	&	2.89	&	37.48	\\
F-43m	&	VH$_2$	&	1.8	&	60.7	&	0.7	&	89.7	&	0.26	&	5.88	&	64.37	\\
Fm-3m	&	ScH	&	1.8	&	34.5	&	2.2	&	65.0	&	0.18	&	3.22	&	58.30	\\
Fm-3m	&	AcH	&	1.7	&	95.7	&	2.2	&	91.8	&	0.11	&	5.17	&	51.14	\\
Pm-3n	&	Ru$_2$H$_6$	&	1.7	&	129.2	&	13.8	&	97.2	&	0.27	&	7.08	&	58.62	\\
Pm-3m	&	TiH$_3$	&	1.1	&	101.3	&	6.1	&	85.6	&	0.29	&	4.58	&	67.37	\\
Pm-3m	&	BaH	&	1.0	&	72.3	&	3.0	&	80.4	&	0.11	&	3.24	&	44.17	\\
Fm-3m	&	OsH	&	1.0	&	61.1	&	1.6	&	90.8	&	0.16	&	3.91	&	40.45	\\
Fm-3m	&	CoH$_2$	&	0.6	&	78.1	&	3.1	&	78.7	&	0.29	&	6.39	&	64.75	\\
Pm-3m	&	ScH$_3$	&	0.6	&	67.8	&	13.8	&	0.0	&	0.27	&	0.52	&	68.36	\\
Fm-3m	&	RaH	&	0.4	&	37.7	&	6.6	&	89.2	&	0.11	&	4.11	&	57.34	\\
F-43m	&	NiH$_2$	&	0.4	&	105.7	&	18.7	&	1.9	&	0.29	&	0.84	&	59.53	\\
Fm-3m	&	RhH	&	0.4	&	40.9	&	5.5	&	100.0	&	0.17	&	7.81	&	41.11	\\
Fm-3m	&	RuH	&	0.3	&	45.6	&	3.0	&	94.6	&	0.18	&	4.87	&	44.87	\\
Fm-3m	&	YH$_3$	&	0.3	&	340.2	&	7.7	&	32.1	&	0.22	&	1.66	&	60.66	\\
Pm-3m	&	KH	&	0.3	&	593.9	&	33.6	&	4.2	&	0.16	&	0.34	&	55.23	\\
Fm-3m	&	YH	&	0.2	&	29.0	&	2.5	&	58.4	&	0.14	&	2.36	&	58.11	\\
F-43m	&	CoH$_2$	&	0.2	&	73.6	&	2.5	&	80.2	&	0.30	&	6.40	&	61.23	\\
Pm-3n	&	Sc$_2$H$_6$	&	0.1	&	322.4	&	31.9	&	41.9	&	0.27	&	2.42	&	61.48	\\
Fm-3m	&	FeH	&	0.0	&	38.4	&	1.5	&	97.0	&	0.23	&	8.29	&	51.58	\\
Fm-3m	&	PdH	&	0.0	&	90.3	&	9.7	&	74.9	&	0.17	&	2.24	&	43.82	\\
Fm-3m	&	CsH$_3$	&	0.0	&	166.9	&	47.1	&	25.9	&	0.20	&	1.04	&	53.94	\\
Fm-3m	&	CsH	&	0.0	&	59.8	&	26.8	&	66.9	&	0.12	&	2.04	&	52.76	\\
Pm-3n	&	Fe$_2$H$_6$	&	0.0	&	91.6	&	11.4	&	94.8	&	0.58	&	10.30	&	63.31	\\
Fm-3m	&	AgH	&	0.0	&	77.7	&	21.9	&	7.6	&	0.16	&	0.37	&	46.62	\\
Fm-3m	&	ScH$_3$	&	0.0	&	243.8	&	3.3	&	46.9	&	0.27	&	2.63	&	61.17	\\
Fm-3m	&	NiH$_2$	&	0.0	&	79.6	&	20.0	&	4.7	&	0.28	&	1.10	&	63.16	\\
Fm-3m	&	FrH	&	0.0	&	79.2	&	28.6	&	52.8	&	0.11	&	1.93	&	56.72	\\
Fm-3m	&	CuH	&	0.0	&	76.5	&	13.0	&	8.3	&	0.22	&	0.42	&	51.11	\\
Fm-3m	&	NiH	&	0.0	&	59.9	&	4.2	&	77.9	&	0.23	&	4.00	&	50.34	\\
F-43m	&	CaH$_2$	&	0.0	&	46.2	&	6.3	&	36.1	&	0.21	&	0.87	&	62.69	\\
F-43m	&	SrH$_2$	&	0.0	&	33.4	&	7.5	&	1.1	&	0.18	&	0.19	&	62.49	\\
Fm-3m	&	CoH	&	0.0	&	32.2	&	2.2	&	100.0	&	0.24	&	10.95	&	48.81	\\
F-43m	&	RaH$_2$	&	0.0	&	24.0	&	2.2	&	0.0	&	0.15	&	0.21	&	63.49	\\
F-43m	&	BaH$_2$	&	0.0	&	9.1	&	0.3	&	0.0	&	0.15	&	0.05	&	59.44	\\

\hline
\end{longtable}

\section*{Appendix 7. USPEX Multiobjective (Pareto) Optimization}

USPEX implements Pareto optimization for the simultaneous treatment of multiple objectives \cite{allahyari2019multi}. In the present calculations, the USPEX workflow used an enthalpy-based objective together with a custom electronic objective. After each structural relaxation, a Python post-processing script read the VASP output, calculated $\alpha\beta\rho$, and wrote a dummy \texttt{EIGENVAL} file in which the nominal band-gap field was proportional to $\alpha\beta\rho$. USPEX then read this field as its second optimization objective.

The enthalpic competitiveness used in the analysis is expressed as $\Delta E=E-E_0$, where $E_0$ is the convex-hull enthalpy at the relevant composition. Thus, the electronic objective supplied to USPEX represents $\alpha\beta\rho$, while the reported enthalpic criterion is evaluated relative to the convex hull. This implementation enabled the descriptor to be incorporated into the Pareto search without modifying the USPEX source code.

\begin{figure}[ht]
\centering
\begin{overpic}[width=\textwidth]{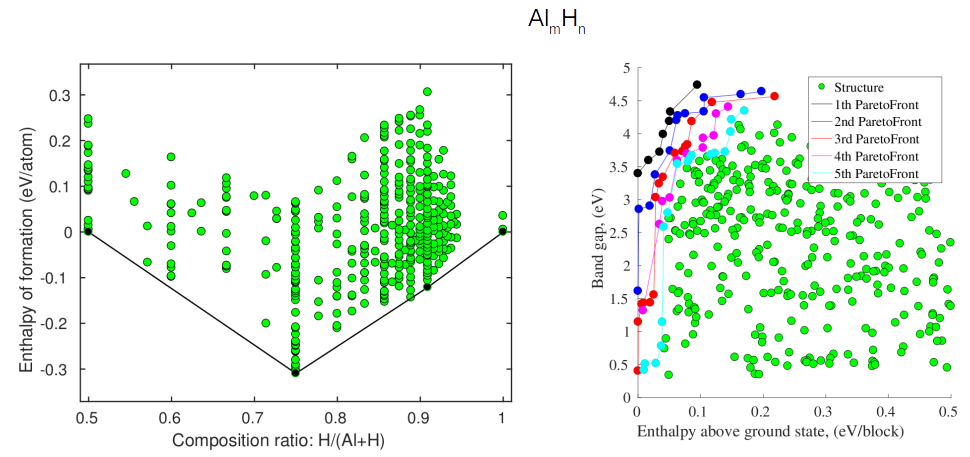}
\put (5,42) {\large\textbf A}
\put (65,42) {\large\textbf B}
\end{overpic}

\caption{(A) Convex-hull diagram for $X_m$H$_n$ at 200~GPa. The black lines define the convex hull. (B) Pareto-front diagram for Al$_m$H$_n$ at 200~GPa. The first, second, third, and fourth Pareto fronts are shown in black, blue, pink, and cyan, respectively.}
\label{fig-Pareto-Al}
\end{figure}

\section*{Appendix 8. Complete Lists of Three-Stage Search Candidates}

The complete S3 candidate lists at 200 and 50~GPa are given below, including entries below the $T_c\geq50$~K cutoff used for the main-text candidate tables and structures treated as duplicates in the main-text presentation. Phonon, EPC, and harmonic Allen--Dynes $T_c$ calculations used a $\mathbf{k}$-point spacing of $0.05$~\AA$^{-1}$ and a $2\times2\times2$ $\mathbf{q}$-point mesh. These values are coarse screening estimates.

\subsection*{Candidates at 200~GPa}

\begin{longtable}{@{\extracolsep{\fill}} c c c c c c c c c c @{}}
\hline
ID &Space& System & $T_c$ & $A_1$ & $\alpha$& $\beta$& $\rho$&$N_f$(states/&$\omega_{max}$\\
&Group &  & (K) & (THz$^2$) & (\%) & (\%) &($\text{\AA}^{-3}$) &Ry/spin/) & (THz)\\
& &  & & & & & &unit cell) &\\
\hline
Mg-2084	&	Fmmm	&	MgH$_{6}$	&	306	&	2081	&	75.4	&	87.3	&	0.42	&	2.829	&	75.1	\\
Mg-55	&	C2/m	&	MgH$_{6}$	&	304	&	1776	&	75.3	&	92.2	&	0.419	&	2.829	&	75.1	\\
Mg-86	&	C2/m	&	MgH$_{6}$	&	297	&	1675	&	75.3	&	96.8	&	0.419	&	2.829	&	75	\\
Mg-3261	&	Im-3m	&	MgH$_{6}$	&	292	&	1729	&	75.3	&	88.4	&	0.419	&	2.83	&	75.1	\\
Mg-53	&	I4/mmm	&	MgH$_{6}$	&	292	&	1599	&	75.3	&	84.9	&	0.419	&	2.829	&	75	\\
Na-1859	&	P-1	&	NaH$_{6}$	&	273	&	1760	&	95.2	&	72.5	&	0.413	&	1.929	&	112.7	\\
Na-2784	&	P2/m	&	NaH$_{6}$	&	272	&	1760	&	95.2	&	91.6	&	0.412	&	1.927	&	112.7	\\
Ca-2691	&	P-1	&	CaH$_{6}$	&	237	&	1162	&	64.8	&	84.3	&	0.361	&	2.697	&	63.2	\\
Ca-281	&	Im-3m	&	CaH$_{6}$	&	237	&	1414	&	64.8	&	86.3	&	0.361	&	2.695	&	63.2	\\
Ca-285	&	R-3m	&	CaH$_{6}$	&	237	&	1264	&	64.8	&	81.7	&	0.361	&	2.695	&	63	\\
Ca-442	&	P-1	&	CaH$_{6}$	&	237	&	1163	&	64.8	&	81.9	&	0.361	&	2.698	&	63.2	\\
K-1818	&	C2/m	&	KH$_{14}$	&	165	&	1525	&	90.5	&	87.1	&	0.455	&	3.207	&	117	\\
K-1974	&	Immm	&	KH$_{12}$	&	163	&	1029	&	81.8	&	89.8	&	0.432	&	2.625	&	121.5	\\
Li-2376	&	P-1	&	Li$_{2}$H$_{6}$	&	158	&	1119	&	77.8	&	80	&	0.447	&	1.689	&	104.2	\\
Al-3279	&	P-1	&	AlH$_{8}$	&	157	&	1039	&	86.7	&	87.3	&	0.451	&	2.704	&	85.6	\\
Li-3030	&	P-1	&	Li$_{2}$H$_{6}$	&	157	&	1112	&	77.8	&	84.5	&	0.451	&	1.686	&	104.3	\\
Al-1686	&	P-1	&	AlH$_{8}$	&	152	&	1036	&	86.7	&	86.1	&	0.45	&	2.701	&	85.7	\\
Al-1378	&	C2/m	&	AlH$_{8}$	&	151	&	1040	&	86.7	&	86.6	&	0.451	&	2.701	&	85.1	\\
Al-2668	&	C2/m	&	AlH$_{8}$	&	151	&	1044	&	86.6	&	92	&	0.448	&	2.709	&	85.2	\\
Li-1674	&	P1	&	Li$_{2}$H$_{6}$	&	151	&	1084	&	77.8	&	83.6	&	0.443	&	1.678	&	104.9	\\
K-2189	&	P-1	&	KH$_{12}$	&	143	&	1028	&	81.9	&	85.8	&	0.432	&	2.621	&	122.6	\\
S-2971	&	P-1	&	S$_{2}$H$_{6}$	&	133	&	696	&	41.7	&	96.4	&	0.309	&	5.325	&	64.9	\\
P-173	&	P-1	&	P$_{2}$H$_{6}$	&	127	&	700	&	48.8	&	92.5	&	0.314	&	5.545	&	88.3	\\
Mg-318	&	I4/mmm	&	MgH$_{4}$	&	113	&	1040	&	86.7	&	86.9	&	0.352	&	1.214	&	116.1	\\
S-2393	&	Im-3m	&	SH$_{3}$	&	108	&	475	&	42.7	&	85	&	0.305	&	2.68	&	52.5	\\
S-3132	&	Im-3m	&	SH$_{3}$	&	108	&	475	&	42.7	&	87.2	&	0.305	&	2.68	&	52.5	\\
S-3183	&	Im-3m	&	SH$_{3}$	&	108	&	475	&	42.7	&	81.2	&	0.305	&	2.68	&	52.6	\\
Li-3240	&	P-1	&	Li$_{2}$H$_{16}$	&	107	&	1004	&	87.8	&	93.2	&	0.534	&	3.389	&	109.1	\\
Ca-289	&	P-1	&	CaH$_{8}$	&	101	&	713	&	53.3	&	76	&	0.394	&	2.924	&	85.6	\\
Si-1665	&	C2/m	&	SiH$_{3}$	&	100	&	649	&	45.3	&	86.1	&	0.324	&	2.585	&	59.9	\\
Si-2059	&	Cm	&	SiH$_{3}$	&	100	&	649	&	45.3	&	86.9	&	0.324	&	2.585	&	59.9	\\
Si-2153	&	R-3m	&	SiH$_{3}$	&	100	&	500	&	45.3	&	77.8	&	0.324	&	2.585	&	59.8	\\
Si-1427	&	C2/m	&	SiH$_{3}$	&	99	&	649	&	45.3	&	87.3	&	0.324	&	2.585	&	60	\\
Si-1912	&	C2/m	&	SiH$_{3}$	&	99	&	649	&	45.3	&	89	&	0.324	&	2.585	&	59.9	\\
Si-2398	&	Cmmm	&	SiH$_{3}$	&	99	&	649	&	45.3	&	97.8	&	0.324	&	2.585	&	59.9	\\
Si-255	&	P-1	&	Si$_{2}$H$_{6}$	&	99	&	725	&	52.5	&	100	&	0.319	&	4.756	&	70.1	\\
Si-1302	&	Amm2	&	SiH$_{3}$	&	98	&	589	&	45.3	&	98.3	&	0.324	&	2.585	&	60.1	\\
Si-783	&	R-3m	&	SiH$_{3}$	&	98	&	501	&	45.3	&	74.3	&	0.324	&	2.585	&	59.8	\\
Si-1141	&	Amm2	&	SiH$_{3}$	&	97	&	589	&	45.3	&	98.3	&	0.324	&	2.585	&	60.3	\\
Si-1899	&	Pm-3m	&	SiH$_{3}$	&	97	&	500	&	45.4	&	90.7	&	0.324	&	2.585	&	59.8	\\
Si-2064	&	Pm-3m	&	SiH$_{3}$	&	97	&	502	&	45.4	&	90.7	&	0.324	&	2.585	&	59.8	\\
Si-2990	&	Pm-3m	&	SiH$_{3}$	&	97	&	502	&	45.3	&	90.7	&	0.324	&	2.585	&	59.8	\\
Si-682	&	Pm-3m	&	SiH$_{3}$	&	97	&	502	&	45.4	&	90.7	&	0.324	&	2.585	&	59.8	\\
Mg-475	&	P4/nmm	&	Mg$_{2}$H$_{6}$	&	96	&	614	&	87.1	&	94.9	&	0.318	&	3.641	&	64.2	\\
Mg-1808	&	Pmmn	&	Mg$_{2}$H$_{6}$	&	95	&	618	&	87.1	&	100	&	0.318	&	3.641	&	63.9	\\
Mg-153	&	Pmmn	&	Mg$_{2}$H$_{6}$	&	94	&	614	&	87.1	&	100	&	0.318	&	3.639	&	63.9	\\
Mg-239	&	P-1	&	Mg$_{2}$H$_{6}$	&	92	&	617	&	87.1	&	91.2	&	0.318	&	3.641	&	63.9	\\
Mg-242	&	P-1	&	Mg$_{2}$H$_{6}$	&	92	&	616	&	87.1	&	91.7	&	0.318	&	3.641	&	63.9	\\
Mg-429	&	P-1	&	Mg$_{2}$H$_{6}$	&	92	&	616	&	87.1	&	93	&	0.318	&	3.641	&	63.9	\\
Na-1654	&	P4/mmm	&	NaH$_{5}$	&	87	&	909	&	97.5	&	98	&	0.391	&	1.294	&	117.3	\\
Ne-1790	&	I4	&	NeH$_{8}$	&	83	&	1027	&	93.7	&	41.8	&	0.46	&	0.347	&	146.5	\\
Cl-3280	&	Cm	&	ClH$_{5}$	&	77	&	1320	&	39.8	&	74.6	&	0.347	&	0.411	&	126.3	\\
Li-632	&	C2	&	LiH$_{10}$	&	77	&	930	&	93.4	&	91.4	&	0.529	&	1.734	&	111.8	\\
Mg-3048	&	P-1	&	MgH$_{8}$	&	76	&	886	&	78.4	&	83.3	&	0.443	&	2.197	&	87.1	\\
Na-1522	&	Cm	&	NaH$_{9}$	&	65	&	683	&	97.9	&	89.6	&	0.452	&	0.82	&	135.3	\\
Cl-1770	&	P6/mmm	&	ClH$_{4}$	&	61	&	656	&	35.8	&	100	&	0.329	&	3.243	&	106.2	\\
Cl-2907	&	P-1	&	ClH$_{4}$	&	61	&	621	&	37.3	&	89.4	&	0.329	&	3.274	&	103.4	\\
Mg-808	&	P-1	&	MgH$_{10}$	&	59	&	847	&	89.3	&	86.3	&	0.451	&	2.496	&	99.2	\\
Li-1319	&	P-1	&	LiH$_{10}$	&	57	&	905	&	92.7	&	87.2	&	0.528	&	1.693	&	122.3	\\
Mg-1086	&	P-1	&	Mg$_{2}$H$_{10}$	&	53	&	598	&	72.6	&	88.1	&	0.383	&	2.146	&	98.1	\\
Mg-2974	&	I4/mmm	&	MgH$_{2}$	&	44	&	702	&	50.7	&	78.6	&	0.253	&	1.437	&	68.3	\\
Mg-1925	&	Fm-3m	&	MgH$_{2}$	&	38	&	253	&	50.7	&	92.8	&	0.253	&	1.437	&	68.3	\\
Mg-1639	&	P-1	&	MgH$_{12}$	&	27	&	785	&	91.8	&	100	&	0.47	&	2.324	&	105.9	\\
Cl-2771	&	Cm	&	ClH$_{13}$	&	22	&	521	&	65	&	91.5	&	0.459	&	1.089	&	127.9	\\
Mg-3467	&	P-1	&	Mg$_{2}$H$_{8}$	&	13	&	517	&	79	&	84.9	&	0.347	&	1.353	&	98.5	\\
K-647	&	P1	&	K$_{2}$H$_{10}$	&	11	&	493	&	57.9	&	71.6	&	0.322	&	2.527	&	123.9	\\
Ca-251	&	P-1	&	CaH$_{4}$	&	10	&	535	&	49.3	&	90	&	0.296	&	1.385	&	84.7	\\

\hline

\end{longtable}
\label{table-long200GPa}

\subsection*{Candidates at 50~GPa}
\begin{longtable}{@{\extracolsep{\fill}} c c c c c c c c c c @{}}
\hline
ID &Space& System & $T_c$ & $A_1$ & $\alpha$& $\beta$& $\rho$&$N_f$(states/Ry&$\omega_{max}$\\
&Group &  & (K) & (THz$^2$) & (\%) & (\%) &($\text{\AA}^{-3}$) &/spin/unit cell) & (THz)\\
\hline
Na-48	&	Pm-3m	&	NaH$_{6}$	&	236	&	1223	&	80.2	&	85.9	&	0.277	&	2.728	&	91.5	\\
Na-1242	&	P-1	&	NaH$_{6}$	&	234	&	1124	&	79.8	&	96.1	&	0.278	&	2.724	&	92.1	\\
Na-643	&	P2/m	&	NaH$_{6}$	&	224	&	1135	&	80.5	&	95.4	&	0.277	&	2.729	&	91.5	\\
K-2925	&	Immm	&	KH$_{12}$	&	149	&	746	&	86.7	&	77.9	&	0.278	&	3.262	&	112.8	\\
Na-752	&	R-3m	&	NaH$_{6}$	&	110	&	786	&	85.7	&	67.5	&	0.271	&	2.436	&	94.8	\\
Na-1531	&	P-1	&	NaH$_{6}$	&	109	&	799	&	85.1	&	85.9	&	0.273	&	2.455	&	97.9	\\
Na-645	&	C2/m	&	NaH$_{6}$	&	96	&	797	&	85.6	&	91.7	&	0.271	&	2.411	&	96.5	\\
P-347	&	P-1	&	PH$_{8}$	&	76	&	389	&	73.7	&	85.5	&	0.278	&	3.297	&	122.8	\\
Sc-3044	&	C2/m	&	ScH$_{10}$	&	73	&	395	&	45.8	&	71.9	&	0.282	&	3.111	&	128.5	\\
Sc-805	&	C2/m	&	ScH$_{6}$	&	71	&	372	&	36.9	&	72.8	&	0.253	&	3.036	&	101.6	\\
Si-2744	&	C2/m	&	SiH$_{8}$	&	67	&	615	&	79.4	&	90.6	&	0.284	&	1.843	&	92.3	\\
Sc-1688	&	C2/m	&	ScH$_{6}$	&	64	&	382	&	34.4	&	69.7	&	0.252	&	2.935	&	90.7	\\
Sc-858	&	Cmmm	&	ScH$_{6}$	&	59	&	382	&	34.4	&	96.6	&	0.252	&	2.941	&	90.7	\\
Li-2877	&	P-1	&	LiH$_{6}$	&	50	&	665	&	78.4	&	71.7	&	0.333	&	1.756	&	97.5	\\
Li-1555	&	P-1	&	LiH$_{6}$	&	48	&	664	&	78.4	&	73	&	0.333	&	1.762	&	97	\\
P-247	&	P1	&	PH$_{8}$	&	41	&	345	&	74.9	&	79.1	&	0.279	&	3.385	&	119.1	\\
Sc-644	&	C2/m	&	ScH$_{8}$	&	36	&	285	&	32.4	&	55.5	&	0.27	&	2.824	&	106.4	\\
Si-2285	&	P-1	&	Si$_{2}$H$_{11}$	&	35	&	299	&	61.7	&	91.2	&	0.251	&	4.695	&	122.2	\\
Sc-255	&	C2/m	&	ScH$_{8}$	&	27	&	289	&	31.8	&	56.6	&	0.27	&	2.842	&	106.2	\\
Al-2333	&	Cmcm	&	Al$_{2}$H$_{6}$	&	8	&	214	&	59.6	&	79.5	&	0.219	&	2.03	&	57.7	\\

\hline
\end{longtable}
\label{table-long50GPa}

\end{document}